\documentclass{article}
\usepackage[utf8]{inputenc}
\usepackage{cite}
\usepackage{authblk}
\usepackage{hyperref}
\usepackage{amsmath}
\usepackage{leitian_v2}
\usepackage{amssymb}
\usepackage[margin=1.3in]{geometry}
\usepackage{graphicx}
\usepackage{upgreek}

\usepackage{amsfonts}
\usepackage{textcomp}
\usepackage{comment}
\usepackage{array}
\usepackage{multirow}
\usepackage[table]{xcolor}
\usepackage{caption} 
\usepackage{booktabs}
\usepackage[ruled, lined, comments numbered, longend]{algorithm2e}

\title{Local Conditional Neural Fields for Versatile and
Generalizable Large-Scale Reconstructions in Computational Imaging}

\author[1]{Hao Wang}
\author[1]{Jiabei Zhu}
\author[1,$\dagger$]{Yunzhe Li}
\author[1]{Qianwan Yang}
\author[1,2,*]{Lei Tian}

\affil[1]{Department of Electrical and Computer Engineering, Boston University, Boston, MA 02215, USA.}
\affil[2]{Department of Biomedical Engineering, Boston University, Boston, MA 02215, USA.}
\affil[$\dagger$]{Current address: Department of Electrical Engineering \& Computer Sciences, University of California, Berkeley, CA 94720, USA.}
\affil[*]{Correspondence: leitian@bu.edu}

\begin{document}

\maketitle

\clearpage

\begin{abstract}
Deep learning has transformed computational imaging, but traditional pixel-based representations limit their ability to capture continuous, multiscale details of objects. Here we introduce a novel Local Conditional Neural Fields (LCNF) framework, leveraging a continuous implicit neural representation to address this limitation. LCNF enables flexible object representation and facilitates the reconstruction of multiscale information. We demonstrate the capabilities of LCNF in solving the highly ill-posed inverse problem in Fourier ptychographic microscopy (FPM) with multiplexed measurements, achieving robust, scalable, and generalizable large-scale phase retrieval. Unlike traditional neural fields frameworks, LCNF incorporates a local conditional representation that promotes model generalization, learning multiscale information, and efficient processing of large-scale imaging data. By combining an encoder and a decoder conditioned on a learned latent vector, LCNF achieves versatile continuous-domain super-resolution image reconstruction. We demonstrate accurate reconstruction of wide field-of-view, high-resolution phase images using only a few multiplexed measurements. LCNF robustly captures the continuous object priors and eliminates various phase artifacts, even when it is trained on imperfect datasets. The framework exhibits strong generalization, reconstructing diverse objects even with limited training data. Furthermore, LCNF can be trained on a physics simulator using natural images and successfully applied to experimental measurements on biological samples. Our results highlight the potential of LCNF for solving large-scale inverse problems in computational imaging, with broad applicability in various deep-learning-based techniques.
\end{abstract}

\newpage
\section{Introduction}\label{sec1}
Deep learning has revolutionized the field of computational imaging~\cite{barbastathis2019use,volpe2023roadmap}, providing powerful solutions to enhance performance and address various challenges in areas such as phase retrieval~\cite{sinha2017,wang2020,bostan2020deep,nguyen2018deep,xue2019}, digital holography~\cite{rivenson2018phase,rivenson2019deep}, diffraction tomography~\cite{matlock2023multiple, saba2022physics, liu2022recovery}, ghost imaging~\cite{wang2022far}, super-resolution imaging~\cite{rivenson2017deep, Nehme:18, wang2019deep}, lightfield imaging~\cite{wang2021real, wagner2021deep, xue2022deep}, lensless imaging~\cite{yanny2022deep, khan2020flatnet}, and imaging through scattering media~\cite{li2018deep,lyu2019learning,tahir2022adaptive}.  Computational imaging treats the image formation process as a two-step procedure: the object information is first physically encoded in the measurement through the imaging optics, and then the information is computationally reconstructed by solving an inverse problem. The effectiveness of deep learning in computational imaging lies in their ability to capture the underlying imaging model and exploit object priors, enabling robust solutions to ill-posed inverse problems~\cite{barbastathis2019use}.  
However, the most widely used reconstruction methods in computational imaging rely on discrete pixels to represent the objects. For instance, a Convolutional Neural Network (CNN) for computational imaging is typically trained on a fixed pixel or voxel grid~\cite{barbastathis2019use}.  This representation is inherently limited by the resolution and size of the grid and does not capture the \emph{continuous} nature and \emph{multiscale} details of the physical objects. Furthermore, the pixel grid representation poses challenges in scaling to process and store large-scale multi-dimensional computational imaging data~\cite{liu2022recovery}. 

To overcome these limitations, we propose a novel deep learning framework called Local Conditional Neural Fields (LCNF) to solve the imaging inverse problem using a \emph{continuous-domain} implicit neural representation that is both compact and highly generalizable.  By utilizing a continuous representation of objects, the LCNF framework offers a more natural and flexible representation that can capture fine-grained details and reconstruct object features of varying scales. We showcase the unique capabilities of LCNF to solve the highly ill-posed inverse problem in Fourier ptychographic microscopy (FPM) with multiplexed measurements~\cite{Tian:15, tian2014multiplexed}, demonstrating robust, scalable, and generalizable large-scale phase retrieval. 

The Neural Fields (NF) framework~\cite{xie2022neural} has recently gained significant interest in computer vision for its ability to represent and render continuous 3D scenes~\cite{mildenhall2021nerf}.  Unlike traditional CNN structures, NF uses a coordinate-based representation, where spatial coordinates (e.g. $(x,y)$) are mapped to physical values (e.g. $[0,1]$) using a multi-layer perceptron (MLP). This unique characteristic of NF allows for the encoding of objects in a \emph{continuous} representation, decoupled from a discrete grid. It enables on-demand synthesis of any part of the object by simply querying relevant coordinates across arbitrary dimensions and resolutions. 
Several NF-based deep learning techniques have been introduced in computational imaging for solving inverse problems using continuous object representations~\cite{sun2021coil,reed2021dynamic, lozenski2022memory, shen2022nerp, Zhu:22, ren2023high, liu2022recovery, zhong2021cryodrgn, doi:10.1126/sciadv.adg4671}. However, these methods are limited by the high computational cost and limited generalization ability. They either require retraining a new NF network for each object reconstruction~\cite{sun2021coil, reed2021dynamic, lozenski2022memory, shen2022nerp, Zhu:22, ren2023high,liu2022recovery, doi:10.1126/sciadv.adg4671} or suffer from the limited representation power of the latent space learned only on the global scale~\cite{zhong2021cryodrgn}, restricting their ability to generalize to diverse objects. 

Our proposed LCNF framework overcomes these limitations by leveraging a \emph{local conditional} NF representation. The \emph{conditional} representation embeds measurement-specific information into the latent space that promotes model generalization. Additionally, the \emph{local} representation allows for the incorporation of multiscale information and enables efficient processing of large-scale imaging data. Together, LCNF enables highly scalable and generalizable deep learning-based image reconstructions. 

\begin{figure}[t]%
\centering
\includegraphics[width=1\linewidth]{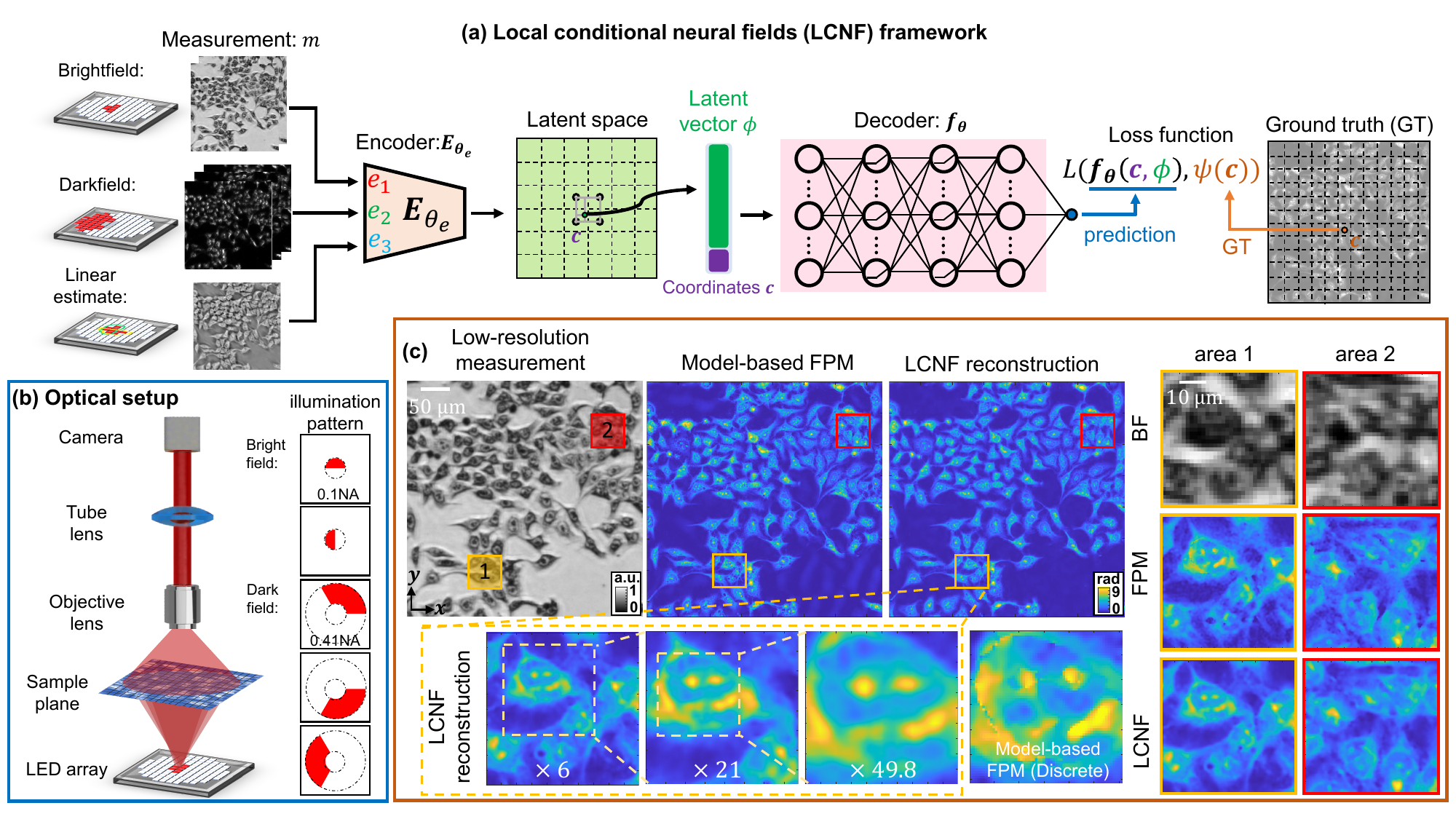}
\caption{Conceptual illustration of our LCNF framework for FPM reconstruction. (a) LCNF employs a CNN encoder to learn measurement-specific information and encode them into a latent-space representation. The MLP decoder reconstructs the phase values at specific locations with an increased spatial resolution by synthesizing local conditional information from the corresponding latent vectors. (b) FPM experimental setup and illumination patterns for acquiring multiplexed BF and DF measurements. (c) Example low-resolution BF measurement, high-resolution phase reconstruction from the model-based FPM algorithm, and our LCNF method. The LCNF learns a generalizable continuous-domain representation and can render phase maps on an arbitrary pixel grid (illustration with $6\times,21\times,49.8\times$ upsampling compared with the raw measurement image).}
 \label{fig1}
\end{figure}

A conceptual illustration of our proposed LCNF framework for FPM reconstruction is shown in Fig.~\ref{fig1}(a). Building on the concept of conditional NF from prior work~\cite{xie2022neural}, our framework utilizes a CNN-based \emph{encoder} to learn measurement-specific information from a set of 2D multiplexed FPM measurements and encode them into a compact latent-space representation. In FPM, the phase information of an object point is spread across multiple pixels on the measured images due to light diffraction. The CNN-based encoder effectively extracts this information by utilizing its extended receptive field, condensing them into latent vectors. Next, an MLP \emph{decoder} is employed to reconstruct the phase values of the object at specific locations based on the corresponding latent information. 
Unlike the traditional NF framework~\cite{mildenhall2021nerf, sun2021coil, reed2021dynamic, lozenski2022memory, Zhu:22, ren2023high, liu2022recovery} that perform a one-to-one mapping between a single coordinate to the corresponding object value, our decoder is conditioned on a learned latent vector that incorporates information across a region of the input images. This conditioning enables adaptation to different objects since each set of measurements is projected onto a distinct latent space representation by the CNN-based encoder. A crucial aspect of FPM reconstruction is achieving ``super-resolution'' reconstruction, surpassing the diffraction limit of the input measurements. To achieve this goal, our framework extracts ``super-resolved'' latent information beyond the ``discrete'' pixel grid in the measurement by incorporating the Local Implicit Image Function (LIIF) method~\cite{chen2021learning} into the decoding process. By combining these components, our LCNF framework achieves versatile deep-learning-based continuous-domain super-resolution image reconstruction based on low-resolution measurements that is applicable to arbitrary objects with varying spatial scales and resolutions.

In this study, we present the capabilities of our proposed LCNF framework for large-scale phase reconstruction based on multiplexed FPM measurements. FPM is a well-established computational imaging technique that combines synthetic aperture and phase retrieval principles to achieve high-resolution reconstructions of amplitude and phase images over a wide field-of-view (FOV) using low-resolution intensity images~\cite{zheng2013wide}. Here, we showcase the effectiveness of our LCNF framework in accurately reconstructing continuous-domain high-resolution phase images over a large FOV using only five multiplexed measurements. Notably, our approach eliminates the need for complex Generative Adversarial Network (GAN) training, as required in previous state-of-the-art approaches~\cite{nguyen2018deep,xue2019}. 

Our results highlight the ability of LCNF to capture the continuous and smooth priors of the object, enabling robust reconstruction of high-resolution phase images. First, using experimental datasets captured on Hela cells fixed in ethanol or formalin, we show that the LCNF network can accurately reconstruct complex cellular and subcellular structures. In addition, we highlight the robustness of the LCNF framework when subjected to imperfect training datasets, benefiting from the implicit continuous priors embedded in our framework. Specifically, LCNF effectively eliminates common artifacts encountered in traditional model-based FPM algorithms, such as residual phase unwrapping errors, noise, and background artifacts, without the need for additional parametric or learned constraints. 

Furthermore, we showcase the strong generalization capabilities of our LCNF framework. Firstly, we demonstrate that LCNF can consistently reconstruct high-resolution phase images even when trained with very limited training data or under different experimental conditions. Remarkably, we achieve high-quality reconstructions even when the network is trained on a \emph{single} imaging data pair. This superior generalization capability is attributed to our NF-based training strategy, which utilizes pixels as training pairs and effectively expands the training data from a single paired image to a diverse set of pixels. Moreover, we demonstrate that LCNF can be trained using purely simulated datasets composed of natural images. We show that the simulation-trained LCNF network generalizes well when applied to experimental biological measurements, consistently reconstructing detailed subcellular structures with minimal artifacts. Finally, we establish that all LCNF networks, regardless of the training strategy, reliably reconstruct high-resolution phase images across a wide FOV.  

In summary, we introduce the LCNF framework as a versatile and generalizable approach for solving highly ill-posed large-scale imaging inverse problems in computational imaging. By leveraging a continuous implicit neural representation, LCNF effectively captures continuous multiscale object information from low-resolution measurements. It provides robust super-resolution reconstruction capabilities, bypassing the limitations of traditional model-based and CNN-based methods that rely on discrete representations. The framework's ability to generalize with very limited training data and its capacity to leverage simulated data further enhance its potential for advancing deep learning-based computational imaging techniques, making it highly attractive for challenging application scenarios where collecting experimental training data is both time-consuming and costly.

\newpage
\section{Results}\label{sec2}
\subsection{The LCNF framework}
\label{LCNF framework}

Our LCNF framework for phase reconstruction from multiplexed FPM measurements is illustrated in Fig.~\ref{fig1}(a). The encoder $E_{\theta_e}$ takes six low-resolution images as input and projects the learned low-dimensional information into a latent space. The input images consist of two brightfield (BF) and three darkfield (DF) intensity measurements captured with the illumination patterns shown in Fig.~\ref{fig1}(b), along with a low-resolution linear estimate of the object's phase computed from the two BF measurements using the differential phase contrast (DPC) method~\cite{Tian:15}. To handle the distinct distributions of BF, DF, and DPC images, three separate encoders $\{e_1,e_2,e_3\}$ are employed to effectively extract the underlying latent information. Each encoder utilizes convolutional layers and residual blocks~\cite{he2016deep} to extract spatial features. The lateral dimensions of the spatial features match those of the input image, allowing for direct coordinate-dependent latent information retrieval during decoding.  The spatial features learned from the three encoders are concatenated to form the final latent space representation $\Phi\in\mathbb{R}^{H\times W\times D}$, where $H$ and $W$ represent the lateral dimensions and $D$ represents the total number of concatenated feature maps in the latent space. 

To enable high-resolution phase reconstruction independent of a fixed grid, a five-layer MLP denoted as $f_{\theta}$ is employed as the decoder. For local conditioning, a specific latent vector $\phi\in\mathbb{R}^{1\times 1\times D}$ is concatenated with the corresponding spatial coordinate $\bc$ before being inputted to $f_{\theta}$. This conditioning mechanism ensures that the learned mapping by the MLP is dependent on the input measurement, allowing for generalizability across different objects.
The output of the decoder is a scalar representing the predicted phase value at the location $\bc$. The LCNF network is trained end-to-end in a supervised manner by minimizing the loss function $L$:
\be
\underset{\theta_e,\theta}{\operatorname{min}} L(f_{\theta}(\bc, \phi), \psi(\bc)),
\label{equation1}
\ee
where $\theta_e$ and $\theta$ represent the network parameters of the encoder and decoder respectively, $\phi=E_{\theta_e}(m,\bc)$ is the latent vector encoded from the input $m$ for the queried coordinate~$\bc$, and $\psi(\bc)$ is the high-resolution ground-truth phase value at the position $\bc$. The ground-truth phase images are reconstructed using separate standard FPM measurements and a previously developed model-based reconstruction algorithm~\cite{tian2014multiplexed, xue2019}. 

A key aspect of FPM is the reconstruction of super-resolved images beyond the low-resolution input. To facilitate the learning of high-resolution information beyond the low-resolution $H\times W$ grid, the LCNF network is also trained on ``off-the-grid''  high-resolution data queried from a denser grid $H'\times W'$. However, the corresponding ``off-the-grid'' latent vector is not readily available from the encoded latent space. In practice, the nearest latent vector (based on the Euclidean distance) is used for the decoder. Additionally, to inform the decoder about the relative position of the queried ``off-the-grid'' location with respect to the nearest latent vector location, the implementation of Eq.~\eqref{equation1} utilizes their relative coordinate $\Delta \bc$ instead of the absolute coordinate $\bc$, following the approach introduced in the LIIF method~\cite{chen2021learning}. Furthermore, to utilize the information provided by the neighboring latent vectors and improve the continuity of the reconstruction, enhancement techniques including feature unfolding, local ensemble, and cell decoding~\cite{chen2021learning} are applied.

After training, the LCNF network allows for querying arbitrary points on the object by providing the corresponding low-dimensional measurements and the queried coordinates as the network input. This eliminates the requirement for a fixed input grid in traditional model-based and deep neural network architectures. The high-resolution phase reconstruction can be visualized on any desired grid. This feature is demonstrated in the results depicted in Fig.~\ref{fig1}(c), where reconstructions are queried at three distinct pixel densities, showcasing smooth transitions across these diverse spatial scales without any artifacts.

More details about the FPM setup, measurements, and model-based reconstructions are provided in Sections~\ref{exp setup} and \ref{model}. Additionally, further details about the LCNF framework, including data acquisition and preprocessing, network structure, reconstruction enhancement techniques, and network training and inference are provided in Sections~\ref{NF}, and Figs.~\ref{s1}, \ref{s3} and \ref{s2}.

\subsection{LCNF reconstruction trained with experimental dataset}
\label{exp_result}
\begin{figure}[t]
\centering
\includegraphics[width=1\linewidth]{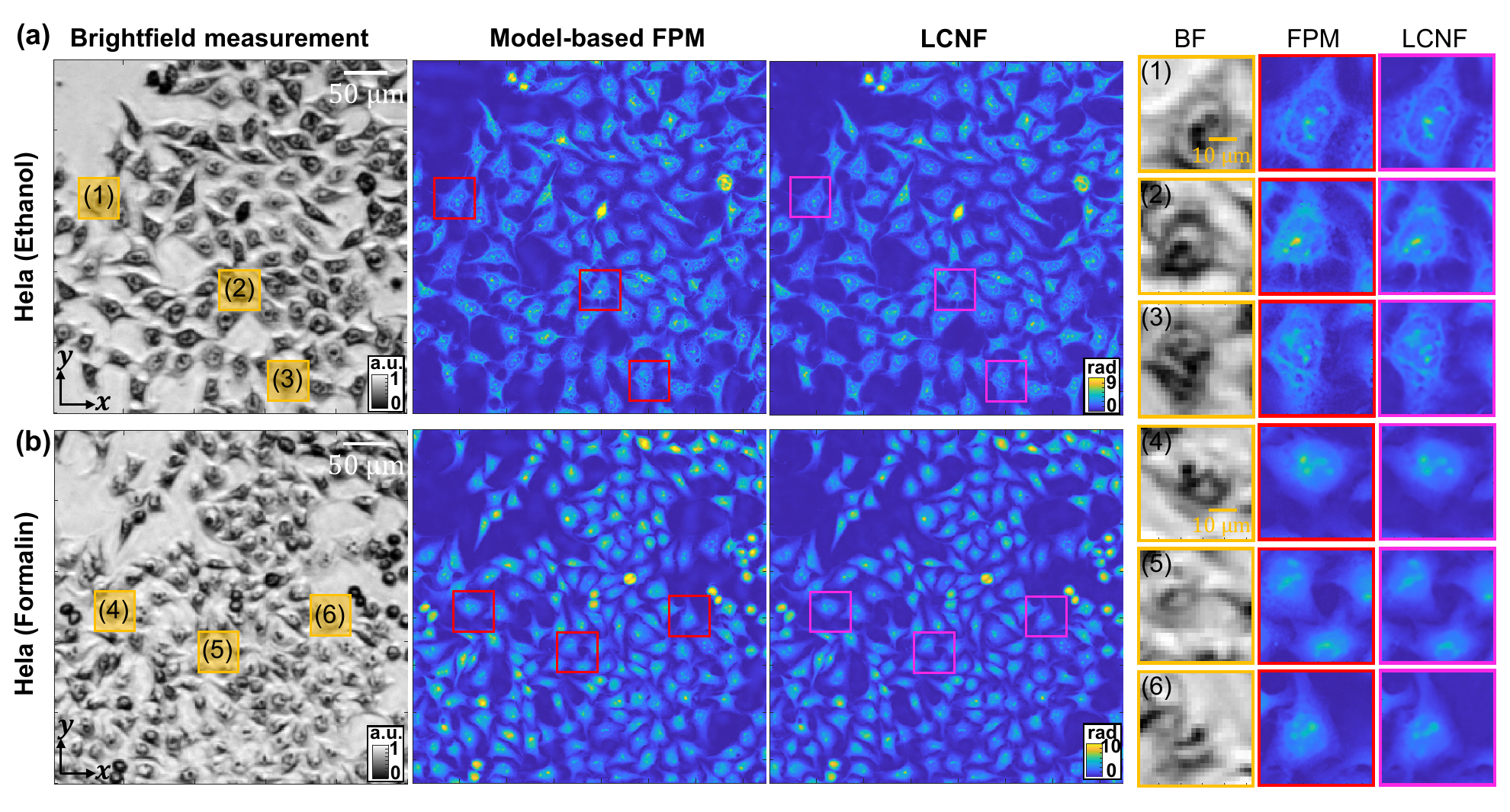}
\caption{Reconstruction results using the LCNF trained with our experimental dataset. An example of BF low-resolution intensity image, the model-based FPM reconstruction, and our LCNF network-based reconstruction for (a) Hela cells fixed with ethanol, and (b) Hela cells fixed with formalin. Subareas (1)-(6) highlight specific regions of interest, demonstrating our network's capacity to accurately reconstruct subcellular high-resolution features without any artifacts.}
\label{fig2}
\end{figure}

We first evaluate the performance of our LCNF network using Hela cells fixed with ethanol or formalin as imaging samples. The network is trained separately for these two cell types (see Section~\ref{net train}), and the reconstruction results using the network trained on the same cell type are shown in Figs.~\ref{fig1}(c) and ~\ref{fig2}.

In Fig.~\ref{fig1}(c), we present the raw low-resolution BF intensity image, the model-based FPM reconstruction, and our LCNF-based reconstruction of ethanol-fixed Hela cells. Furthermore, we display two small subareas (area 1 and area 2). From the figure, it is evident that our network successfully reconstructs high-resolution phase images from low-resolution intensity images, accurately recovering intricate subcellular structures.

To evaluate the continuous object representation capability of our LCNF network, we conduct queries on arbitrary coordinates within subarea 1, as shown at the bottom of Fig.~\ref{fig1}(c). We perform queries at densities of $6\times, 21\times, 49.8\times$ compared to the input low-resolution intensity image. Our network successfully reconstructs the phase at these density grids. For comparison, we also include the model-based FPM reconstruction of the same area. Due to predefined grids, the FPM reconstruction exhibits discrete grid artifacts in the enlarged image.
Furthermore, it may suffer from phase unwrapping artifacts (see Section~\ref{robust_CNF} for more details). For an additional comparison, we present high-resolution phase reconstructions with a grid density of $105.9\times$ in Fig.~\ref{s8}. In contrast, our network provides a continuous object reconstruction without any discrete or other phase artifacts. 

Figure~\ref{fig2}(a) and (b) showcase additional reconstruction results for Hela cells fixed with ethanol and formalin, respectively. As shown in the figures, we successfully reconstruct high-resolution phase images from the low-resolution intensity images, accurately capturing detailed cellular and subcellular structures without any artifacts, as highlighted in the zoom-in regions (1)-(6).

\subsection{Robustness to phase artifacts}
\label{robust_CNF}
Next, we highlight the robustness of our LCNF network to various phase artifacts that arise from practical FPM experiments, including noise, phase unwrapping errors, and artifacts resulting from an imperfect imaging model. 

\begin{figure}[t]
\centering
\includegraphics[width=1\linewidth]{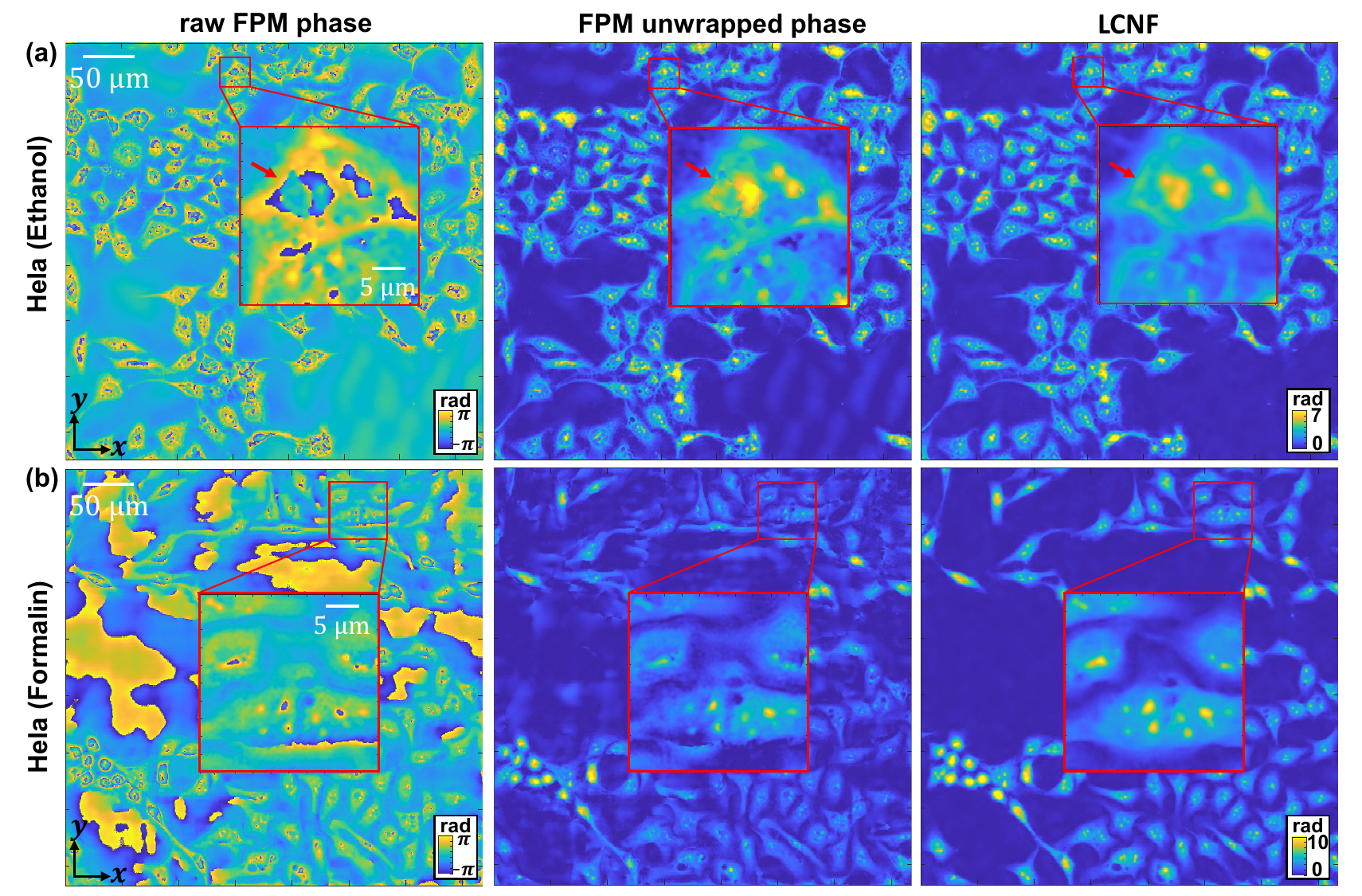}
\caption{LCNF's robustness to phase artifacts. The LCNF reconstruction completely eliminates (a) the discontinuous phase-unwrapping errors (indicated by red arrows) and background ripping artifacts, as well as (b)  random phase noise.}
\label{fig3}
\end{figure}

As shown in Fig.~\ref{fig1}(c) and Fig.~\ref{fig3}(a), the model-based FPM reconstruction may exhibit discontinuous artifacts due to imperfect phase unwrapping when dealing with samples that have a phase range exceeding $2\pi$. Moreover, Fig.~\ref{fig3}(a) illustrates the presence of rippling artifacts in the background region of the model-based FPM reconstruction, possibly resulting from the imperfect FPM imaging model used for the model-based reconstruction~\cite{yeh2015experimental}. Additionally, Fig.~\ref{fig3}(b) demonstrates that the model-based FPM reconstruction can be susceptible to random phase noise. 

In contrast, our LCNF network effectively eliminates these artifacts and achieves accurate, smooth, and continuous reconstructions, \emph{even though it has been trained using imperfect ground-truth images from experiments that inevitably contain these artifacts.} We quantitatively evaluate the artifact-suppression capability of our LCNF network using the method in~\cite{wang:23}. 
Our analysis shows that our LCNF network can reduce the background artifacts by several folds compared with the model-based FPM reconstruction, as illustrated in Fig.~\ref{s4}.

The robustness can be attributed to the implicit continuous priors embedded in our LCNF network structure. The LCNF framework employs a two-step process to achieve continuous representations. Firstly, it encodes the input images into a continuous latent space representation, effectively filtering out noisy information. Secondly, it decodes the queried point by conditioning it on the selected latent vector. This process leverages the continuity priors embedded in the MLP decoder, enabling it to learn a continuous neural representation of the object.

Overall, our network demonstrates robust reconstruction capabilities even when trained with imperfect datasets, benefiting from the continuity of the learned latent space and the continuous representation imposed by the MLP decoder.

\subsection{Generalizability of experimental data trained LCNF network}
\label{general_CNF}

\begin{figure}[t]
\centering
\includegraphics[width=1\linewidth]{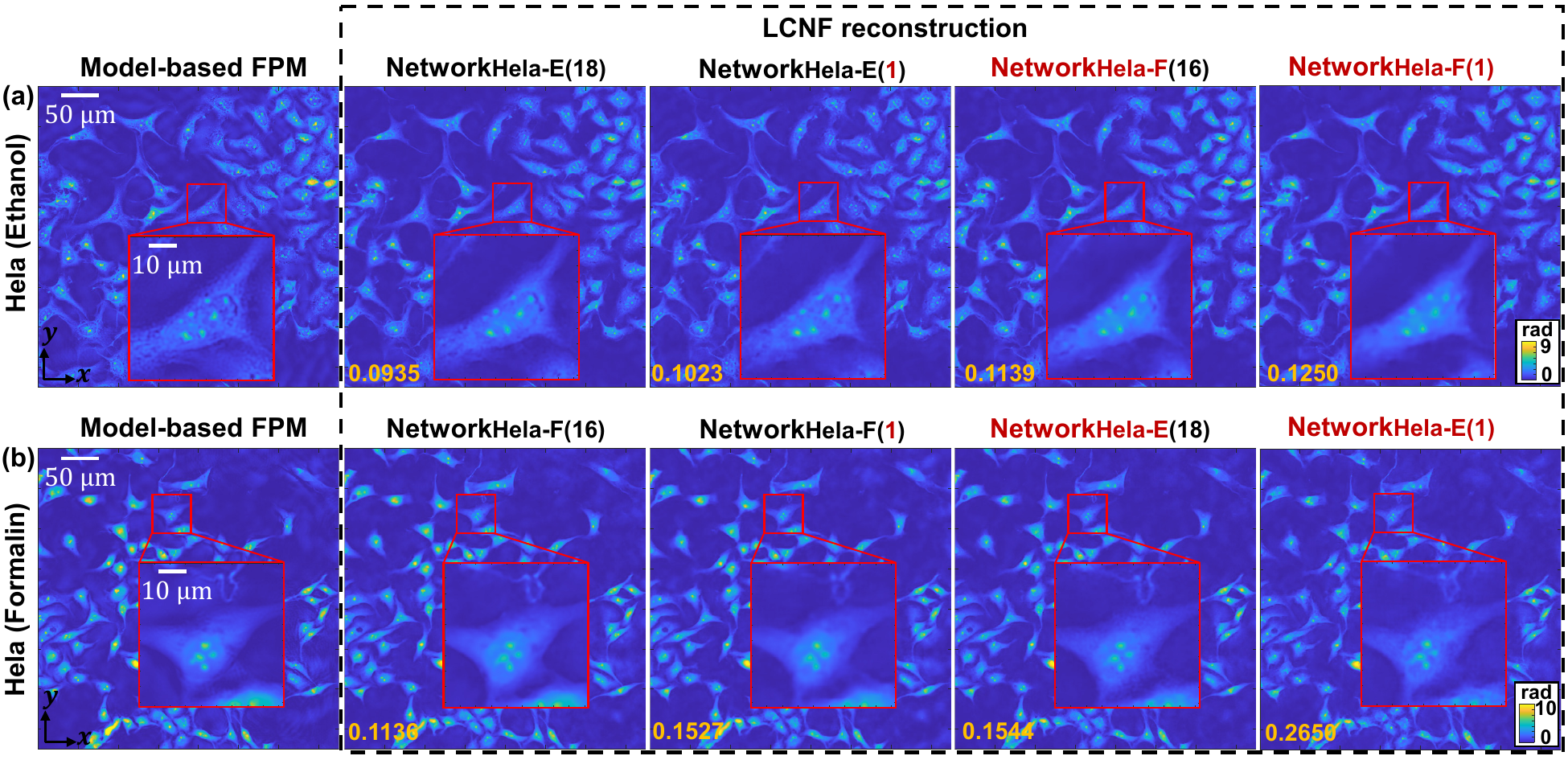}
\caption{Generalization ability of experimental data trained LCNF network. Reconstructions of (a) Hela cells fixed in ethanol and (b) Hela cells fixed in formalin. The MSE of each reconstruction is noted at the bottom of the respective image. Networks used for reconstruction: 
\textbf{Network$_{\text{Hela-E(18)}}$} (trained with 18 pairs of ethanol-fixed Hela cells), \textbf{Network$_{\text{Hela-F(16)}}$} (trained with 16 pairs of formalin-fixed Hela cells), 
\textbf{Network$_{\text{Hela-E(1)}}$} (trained with a single pair of ethanol-fixed Hela cells), and 
\textbf{Network$_{\text{Hela-F(1)}}$} (trained with a single pair of formalin-fixed Hela cells).}
\label{fig4}
\end{figure}

One notable advantage of our LCNF framework is its superior generalization capability, overcoming the limitations of traditional NF frameworks~\cite{liu2022recovery}. To thoroughly evaluate its generalization performance, we conduct training using three distinct types of experimentally collected datasets, as outlined in Section~\ref{net train}. These training scenarios included: (1) utilizing one type of experimental data to train the network and evaluating it on the same type of dataset, (2) training the network with a single pair of data and testing it on the same type of data, and (3) evaluating the aforementioned networks' performance on other types of cells. The LCNF networks consistently demonstrate successful reconstruction of high-resolution phase images across all three training scenarios, as depicted in Fig.~\ref{fig4}.

We quantitatively assess the performance of our LCNF-based reconstructions for both image patches from the FOV regions matching the training conditions and outside the training region. This assessment is informative because spatially varying aberrations are known to degrade FPM reconstructions~\cite{ou2014embedded}. By evaluating the reconstruction quality outside the training FOV, we gain insights into the network's robustness against realistic spatially varying aberrations in our experiment. For the evaluation, we employ the mean square error (MSE) as the objective metric. The results, presented in Fig.~\ref{fig4}, illustrate the robust performance of our LCNF network in all three scenarios, with the corresponding MSE values provided at the bottom of the figure. We also compute the peak signal-to-noise ratio (PSNR), structural similarity index measure (SSIM), and frequency measurement metric (FM). The FM quantifies the recovery of frequency components~\cite{de2013image}, where higher FM values represent the recovery of more frequency components. The quantitative metrics for the results in Fig.~\ref{fig4} are presented in Table~\ref{table1:gen}, and the metrics for an additional 100 image patches outside the training FOV region are provided in Table~\ref{table3:100sample}.

As shown in Fig.~\ref{fig4}, and Tables~\ref{table1:gen} and \ref{table3:100sample}, the MSE generally increases, while the PSNR, SSIM, and FM decrease when training the LCNF network with a very limited dataset or a different type of data compared to the network trained with the same cell type and the full experimental dataset. This indicates that the network's generalization performance generally degrades when it is trained on a smaller training dataset or the distribution of the testing data is shifted from that of the training data, which is expected. However, the changes in the metric values are small and hardly noticeable in the visualizations in Fig.~\ref{fig4}, even for the network trained with \emph{a single} paired training dataset. This highlights the robust generalization performance of our LCNF network. 

In addition, when the network is trained with ethanol-fixed Hela cells  (\textbf{Network$_{\text{Hela-E(18)}}$}) and applied to formalin-fixed Hela cells, the SSIM and FM are slightly higher than those of the network train with formalin-fixed Hela cells (\textbf{Network$_{\text{Hela-F(16)}}$}), as shown in Table~\ref{table3:100sample}. We attribute this ``unusual'' result to the fact that ethanol-fixed Hela cells contain more structural details and provide a broader spectrum compare to formalin-fixed  Hela cells (see Fig.~\ref{fig5}(b)). Therefore, the network trained with ethanol-fixed Hela cells may reconstruct more frequency components and thus yield better results.

We attribute this generalization capability to our novel training strategy, which utilizes \emph{pixels} as the training pairs (as shown in Eq.~\eqref{equation1}). 
By adopting this approach, we effectively expand the training data from a single paired image ($250\times250$ pixels for the input and $1500\times1500$ pixels for the high-resolution reconstruction) to a diverse set of pixels. This enables the network to learn from a larger and more varied dataset, contributing to its remarkable generalization capabilities. As a result, our LCNF network demonstrates the ability to reconstruct high-resolution phase images even when trained with very limited training data. This not only reduces the necessity for a large number of training samples but also expedites the overall experimental process, making it highly suitable for challenging experimental scenarios where collecting experimental training data is both time-consuming and costly.

\subsection{LCNF network generalizes from simulation to experiment}
\label{sim}

\begin{figure}[!t]
\centering
\includegraphics[width=1\linewidth]{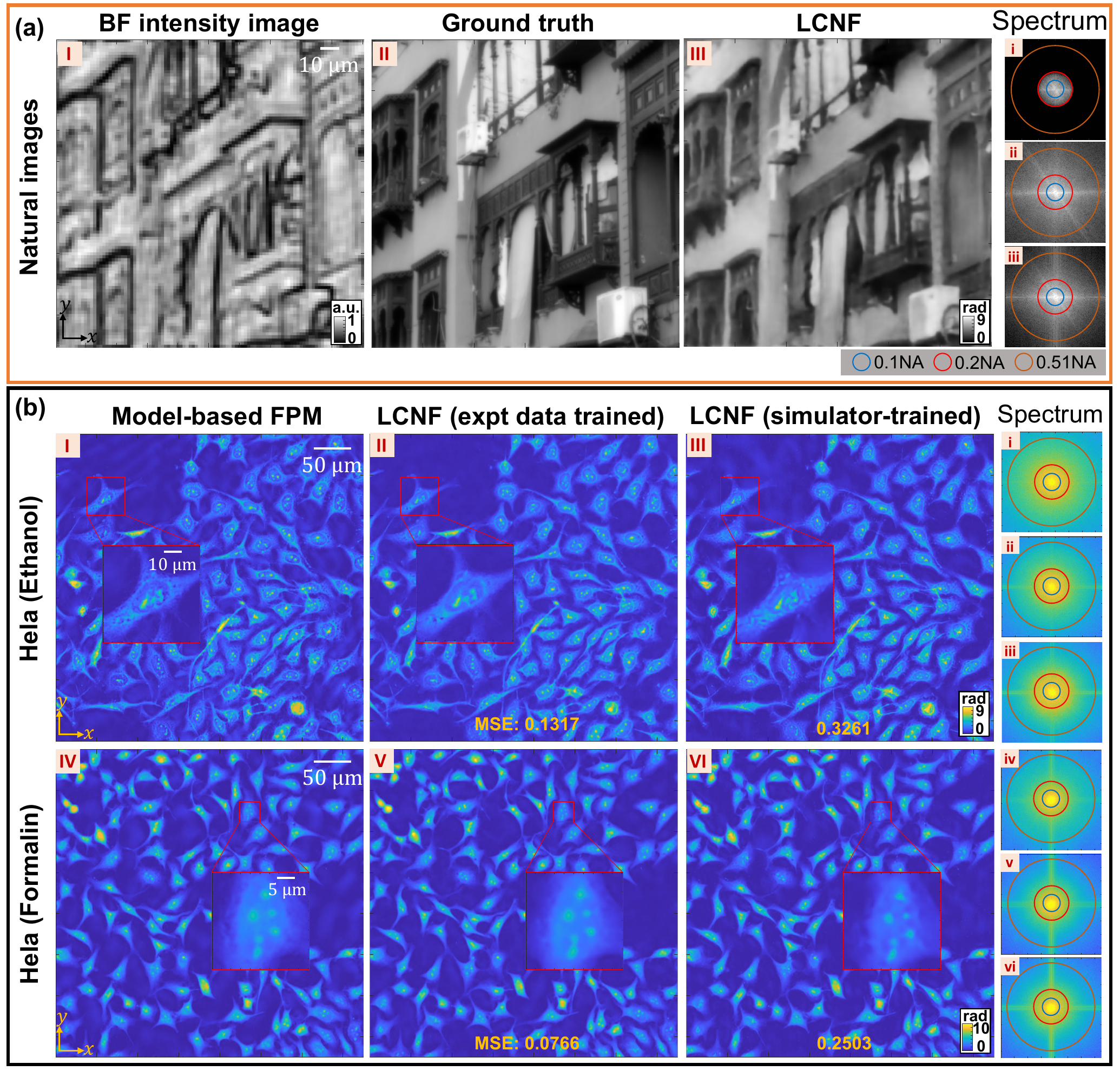}
\caption{Generalization ability of simulation-trained LCNF network. (a) Example simulation on a natural image. I: Simulated multiplexed BF intensity image. II: High-resolution ground-truth image. III: Reconstruction by LCNF. (i-iii) The spectrum of (i) simulated BF intensity image, (ii) ground truth, and (iii) LCNF reconstruction.  Blue, red, and brown circles indicate the bandwidth of the objective lens (0.1 NA), the expected bandwidth of the BF measurement (0.2 NA), and the theoretically achievable reconstruction bandwidth (0.51 NA), respectively. (b) Reconstruction of experimental data from Hela cells fixed in ethanol and formalin using (I, IV) model-based FPM algorithm, (II, V) LCNF network trained with the full experimental dataset with matching cell type (II: \textbf{Network$_{\text{Hela-E(18)}}$}, V: \textbf{Network$_{\text{Hela-F(16)}}$}), and (III, VI) LCNF network trained with simulated dataset (\textbf{Network$_{\text{Simulate}}$}). (i-vi) Spectrum of  reconstructions I-VI labeled by i-vi in the right column, with different circles representing different NAs as in (a).}
\label{fig5}
\end{figure}

We further demonstrate the robust generalization capability of our LCNF network by employing a simulator-trained network for the reconstruction of the experimental Hela cells dataset, as depicted in Fig.~\ref{fig5}. When solving the inverse problem using deep learning-based methods, acquiring paired datasets for network training can be challenging. Traditional NF-based approaches incorporate the imaging forward model within the network, allowing for self-supervised training without the need for paired datasets. However, as mentioned earlier, these methods often lack the ability to generalize across different objects and necessitate separate training for each new object.

An alternative approach is to use the imaging forward model to generate simulated paired datasets for network training~\cite{Wang:19, matlock2023multiple, tahir2022adaptive,xue2022deep,yanny2022deep}. However, in the context of FPM reconstruction, the application of simulator-trained networks has been obstructed by the use of the GAN structure that learns highly specific but less generalizable object priors~\cite{xue2019}. 
Here, we demonstrate straightforward deployment of simulation-based training of our LCNF network and achieve high-resolution, wide-FOV phase reconstructions on the experimental biological dataset (see training details in Section~\ref{preprocess}).

We first evaluate the network's performance on simulated data, as illustrated in Fig.~\ref{fig5}(a).  The results confirm the successful reconstruction of high-resolution phase images from low-resolution intensity images. The MSE, PSNR, SSIM, and FM metrics are presented in Table~\ref{table2:sim}. When evaluating the network using the simulated dataset, we did not apply any preprocessing techniques described in Section~\ref{preprocess} to the ground-truth high-resolution natural images, except for linearly matching pixel values with the phase range of [0, 9]. Consequently, we can assess the network's performance more accurately without the influence of data preprocessing effects. Notably, our network demonstrates high effectiveness in recovering a significant portion of the spatial frequency components, as evidenced in the visualization of the reconstructed spectrum in the right column of Fig.~\ref{fig5}(a) and Table~\ref{table2:sim}. 

Subsequently, we employ our simulator-trained network for the reconstruction of experimentally captured Hela cells datasets, as depicted in Fig.~\ref{fig5}(b). The network successfully reconstructs Hela cells with detailed subcellular structures and recovers the rich spatial frequency components. 

The quantitative metrics, including MSE, PSNR, SSIM and FM, are provided in Table~\ref{table2:sim}. The results show that the simulator-trained network performs slightly worse than the experimental data-trained network in terms of MSE, PSNR, and SSIM.  This is expected because our training data consists only of natural images, which have significantly different image features compared to the cells in the experiment.  However, the FM metrics of our simulation-trained network are consistently higher than all other methods. This observation suggests that training the network on PSD-corrected natural images may promote more effective learning for high-frequency content. To further evaluate the performance of the simulation-trained network, we conduct additional reconstructions on 100 experimental image patches beyond the training FOV. The quantitative metrics are presented in Table~\ref{table3:100sample}.

These results clearly demonstrate the generalization capability of our simulation-trained LCNF network in achieving high-resolution phase reconstructions in the experiment.

\subsection{Robust wide-FOV high-resolution phase reconstruction}
\label{wideFOV}

\begin{figure}[!t]
\centering
\includegraphics[width=1\linewidth]{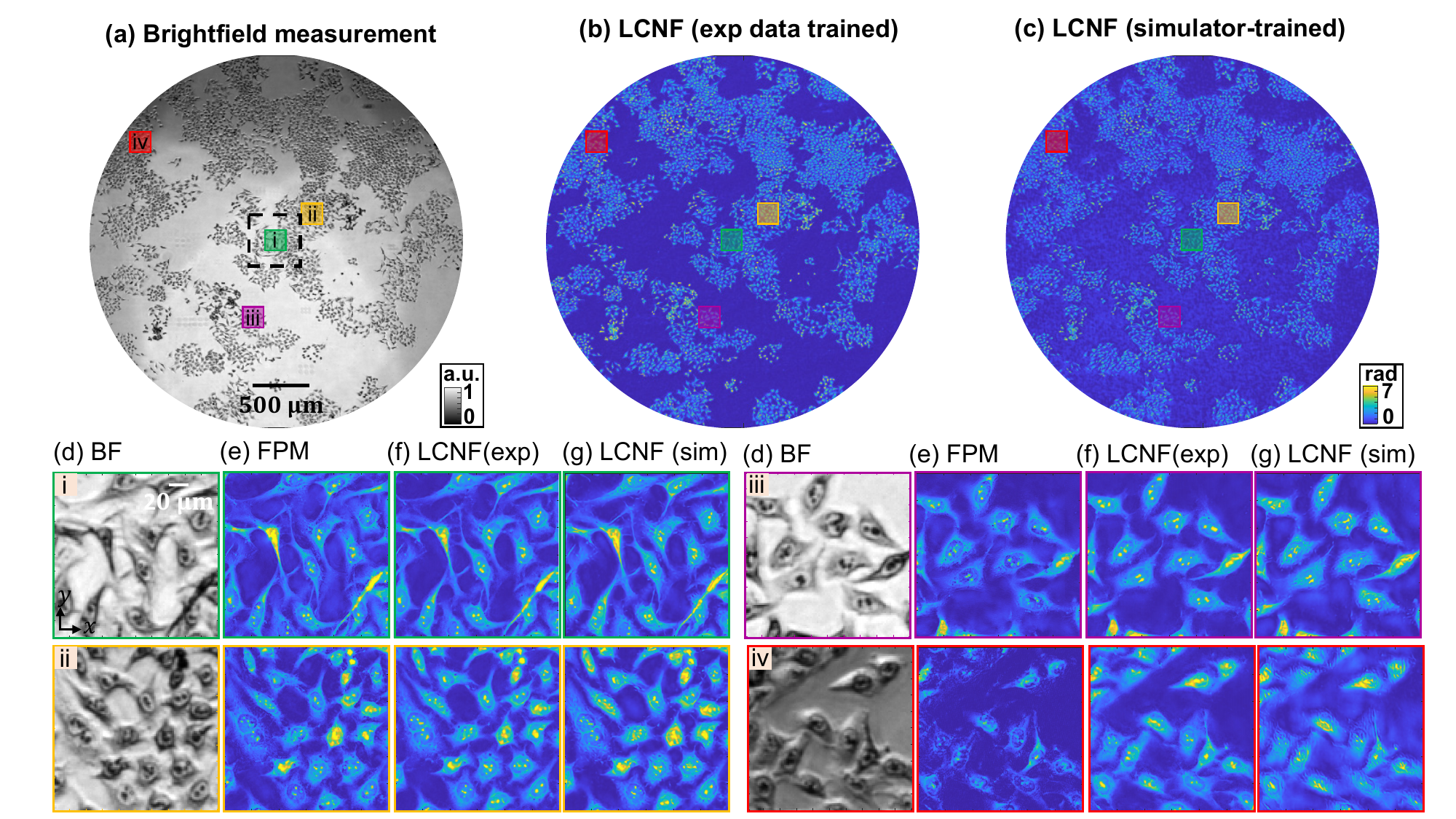}
\caption{Wide-FOV high-resolution phase reconstruction by the LCNF network. (a) BF intensity image captured over a 2160-pixel diameter (3.51 mm) FOV. Wide-FOV reconstruction by training the LCNF network with (b) an experimental dataset and (c) a simulated dataset. (d-g) Selected subareas extracted from the central to the edge of the FOV, identified as i-iv and enclosed within different colored boxes.  (d) BF intensity image. (e) model-based FPM reconstruction. (f) LCNF reconstruction trained with the experimental dataset. The experimental dataset used for training the LCNF network is obtained from the central region, indicated by the dashed black square. (g) LCNF reconstruction trained with the simulated dataset.}
\label{fig6}
\end{figure}

Finally, we employ our LCNF network for wide-FOV high-resolution phase reconstructions, as shown in Fig.~\ref{fig6}. The network is trained solely using the central $250\times250$-pixel region, indicated by the dashed black square in Fig.~\ref{fig6}(a). Subsequently, we perform phase image reconstruction across a much larger FOV that encompasses a circular region with a 2160-pixel diameter in the raw measurements (3.51 mm). The resulting wide-FOV reconstructions (12960-pixel in diameter and 6$\times$ denser pixel grid compared to the input), obtained using \textbf{Network$_{\text{Hela-E(18)}}$} and \textbf{Network$_{\text{Simulate}}$}, are presented in Fig.~\ref{fig6}(b) and (c), respectively. Additionally, Fig.~\ref{fig6}(d)-(g) displays selected subareas extracted from the central to the edge of the full-FOV image. The corresponding images include the BF intensity image, model-based FPM reconstruction, LCNF \textbf{Network$_{\text{Hela-E(18)}}$} reconstruction trained with the experimental dataset (LCNF(exp)), and LCNF \textbf{Network$_{\text{Simulate}}$} reconstruction trained with the simulated dataset (LCNF(sim)). Overall, the LCNF networks achieve high-quality reconstructions, with subcellular features clearly recovered and minimal artifacts. However, at the very edge of the image, as observed in Fig.~\ref{fig6}(d)-(g)iv, some distortions are present.  This behavior is expected due to the spatially varying aberrations present in our microscope setup, which become more pronounced at the edge of the FOV~\cite{ou2014embedded}. The experimental dataset used for training the network is obtained from the central FOV, where aberrations are minimal. The simulation assumes a perfect imaging system without any aberrations. Consequently, the network was not exposed to these aberrations during training. Addressing this limitation will require incorporating a spatially variant imaging model, which we plan to consider in our future work. 

Furthermore, we present additional wide-FOV reconstructions in Figs.~\ref{s5}, \ref{s7}, and \ref{s6} for Hela cells fixed in ethanol or formalin. These reconstructions were obtained using the LCNF networks trained with all the strategies detailed in Section~\ref{general_CNF} and \ref{sim}, based on experimental or simulated datasets. The results further underscore the reliability of our LCNF framework in achieving wide-FOV high-resolution phase reconstructions, regardless of the training strategy employed. Notably, our framework demonstrates excellent performance even when trained with very limited data, including a single paired image at the extreme case, or when utilizing simulated training data.

\section{Discussion and Conclusion}
\label{conclu}

In this study, we have introduced LCNF, a versatile and generalizable deep learning framework for solving large-scale imaging inverse problems. Unlike traditional CNN frameworks, LCNF leverages a continuous implicit neural representation to enable flexible reconstruction of multiscale information.  It introduces a novel local conditioning approach, enhancing its generalization capability compared to existing NF frameworks.  

By applying LCNF to solve the multiplexed FPM reconstruction problem, we demonstrate its effectiveness in achieving continuous-domain super-resolution reconstruction from low-resolution measurements, applicable to objects of varying spatial scales and resolutions. In addition, LCNF exhibits robustness against noisy training data. The LCNF reconstructions are free from common artifacts, such as residual phase unwrapping errors, noise and background ripples, that contaminate the training data obtained by traditional FPM reconstructions. 

Furthermore, LCNF demonstrates remarkable generalization capabilities across different object types and experimental conditions. Notably, LCNF can be effectively trained even with limited datasets, including a single paired image dataset, considerably simplifying the experimental training data collection process. Additionally, we show that LCNF can be entirely on simulated data and generalize well to experimental data without requiring network retraining or transfer learning. This further highlights the robustness and adaptability of our LCNF approach.

LCNF's efficient processing of multiscale information makes it highly suitable for large-FOV high-resolution image reconstruction applications.  We showcase LCNF's ability to robustly perform large-scale super-resolution phase reconstructions using multiplexed FPM measurements, regardless of the training strategy employed.

In summary, we present LCNF as a robust and scalable deep-learning-based continuous-domain reconstruction framework. Its ability to handle large FOV and high-resolution imaging reconstruction, combined with its strong generalization capabilities, makes it suitable for a wide range of computational imaging techniques.

\section{Methods}
\subsection{FPM experimental setup}
\label{exp setup}

Our LCNF network was developed based on the experimental data obtained from our previous study in~\cite{xue2019}. To briefly describe the FPM setup and the data acquisition method, the illumination multiplexing scheme combined patterns used in DPC~\cite{Tian:15} and randomly multiplexed FPM~\cite{tian2014multiplexed,tian2015computational} to efficiently encode high-resolution phase information across a wide FOV. Specifically, we used five LED illumination patterns (central wavelength $630 \mr{nm}$), including two BF semi-circle patterns and three $120^{\circ}$-arc patterns, as illustrated in Fig.~\ref{fig1}(b). To capture the standard FPM dataset, sequential illumination with 185 LEDs was employed. In both illumination schemes, the maximum illumination NA was 0.41. The samples used for testing were unstained Hela cells fixed with ethanol or formalin. Intensity images were collected using a $4\times$, 0.1 NA objective lens (Nikon CFI Plan Achromat) and an sCMOS camera (PCO: pco.edge 5.5) with $2560\times2160$ pixels and a pixel size of $6.5\upmu \mr{m}$.

\subsection{Model-based reconstructions}
\label{model}

\subsubsection{DPC-based phase imaging}
\label{dpc}
Here, we briefly explain the principle of DPC phase imaging; additional details can be found in~\cite{Tian:15}. DPC is a technique used to recover quantitative phase information from intensity images acquired with asymmetric illumination patterns. It offers an improved lateral resolution of $2\times$ compared to the native objective NA.

Under the weak object assumption: $o(\bc)=e^{-\mu(\bc)+i\psi(\bc)}\approx 1-\mu(\bc)+i\psi(\bc)$, where $\mu(\bc)$ represents absorption and $\psi(\bc)$ represents phase, a BF intensity measurement $I_s(\bc)$ can be approximated to have a linear relationship with the sample~\cite{Tian:15}:
\be
I_S(\bu)=B\delta(\bu) + H_{abs}(\bu)M(\bu) + H_{ph}\Psi(\bu),
\label{equ3}
\ee
where $I_S(\bu), M(\bu), \Psi(\bu)$ denote the spectrum of $I_s(\bc), \mu(\bc)$ and $\psi(\bc)$, respectively, and $\bu=(u_{x},u_{y})$ represents the spatial frequency. $B$ is a constant representing the background signal, and $\delta(\bu)$ is the Dirac delta function. $H_{abs}, H_{ph}$ are the transfer functions for amplitude and phase, respectively~\cite{Tian:15}.

By subtracting the background term and normalizing the acquired BF intensity image, the DPC reconstruction can be formulated as:
\be
\underset{M,\Psi} {\operatorname{min}}\sum_{j=1}^{N_{\mr{bf}}}\|I_{S-,j}(\bu) - H_{abs,j}(\bu)M(\bu) - H_{ph,j}\Psi(\bu)\|_2^2 + \tau_1 R_1(M(\bu)) + \tau_2R_2(\Psi(\bu)),
\label{equ4}
\ee
where $I_{S-}(\bu)$ represents the spectrum of the background-subtracted intensity image, $j$ is the index of DPC measurements, $N_{\mr{bf}}=2$ denotes the number of captured BF images, and $\|\cdot\|_2$ represents the $L_2$ norm. $\tau_1, \tau_2$ are the regularization parameters, and $R_1$ and $R_2$ represent the regularization terms that incorporate prior information about the sample. Here, we utilized the $L_2$ regularization to solve inverse problem~\cite{Tian:15}. 

It should be noted that DPC reconstruction relies on the weak object approximation, which means it only provides accurate results when the phase change of the sample is below 0.64 radians~\cite{chen2018quantitative}. However, in our experiment, the Hela cell samples are fixed in ethanol or formalin, causing phase changes exceeding $2\pi$. This violates the weak object approximation and leads to an underestimation of the object's phase.  Despite its limitations, the DPC estimation serves as a useful low-resolution initial guess for the object's phase~\cite{tian2015computational,wang:23}. Therefore, we input this estimation into our network.

\subsubsection{FPM forward model}
The forward model of FPM describes the intensity image obtained from a single LED illumination.  After appropriate normalization to account for the system magnification,  it can be expressed as:
\be
    I_i(\mr{\bc}) = |\mathcal{F}^{-1}_{[O(\mr{\bu}-\mr{\bu}_i)P(\bu)]}(\bc)|^2,
    \label{equ1}
\ee
where $I_i(\bc)$ represents the captured low-resolution intensity image for the $i^{\mr{th}}$ LED. $|\cdot|$ takes the amplitude of the complex field, and $\bc=(x, y)$ denotes the lateral coordinates. $\mathcal{F}^{-1}$ represents the inverse Fourier transform, and $O(\bu)$ is the spectrum of the object $o(\bc)$.
Each LED illumination is modeled as a plane wave with spatial frequency $\bu_i=(u_{xi},u_{yi})=({\sin\theta_{xi}}/{\lambda}, {\sin\theta_{yi}}/{\lambda})$, where $(\theta_{xi}, \theta_{yi})$ defines the illumination angle of the $i^{\mr{th}}$ LED and $\lambda$ denotes the central wavelength. The pupil function of the microscope, denoted by $P(\bu)$, is a circular low-pass filter with a diameter of $2\mr{NA}/\lambda$, set by the objective lens NA. 

In the case of multiplexed illumination, the sample is illuminated by different sets of LEDs based on different illumination patterns, as depicted in Fig.~\ref{fig1}(b). The captured intensity image can be modeled as the sum of multiple intensity images obtained from individual LEDs~\cite{tian2014multiplexed}:
\be
    I_S(\mr{\bc})=\sum_{i\in S}I_i(\bc),
\label{equ2}
\ee
where the symbol $\in$ indicates that $i$ is an element of the illumination set $S$.

\subsubsection{Model-based FPM reconstruction}
\label{fpm}
FPM is a recently developed computational imaging technique that enables increasing the imaging system's space-bandwidth product (SBP) by synthesizing multiple low-resolution images into a high-resolution image across a wide FOV~\cite{zheng2013wide}. 
The FPM reconstruction involves solving a non-convex optimization problem that jointly estimates the object $O(\bu)$ and the  pupil function $P(\bu)$ by solving a minimization problem:
\be
\underset{O(\bk),P(\bk),\{b_i\}}{\operatorname{min}}\sum_{i=1}^{N_\mathrm{led}}
\big\| \sqrt{I_i(\bc)}-|\mathcal{F}^{-1}_{[O(\mr{\bu}-\mr{\bu_i})P(\bu)]}(\bc)| - b_i\big\|^2_2,
\label{equ5}
\ee
where $b_i$ is the background offset for the $i^{\mr{th}}$ image, and $N_\mathrm{led}$ is the total number of LEDs used in the sequential FPM measurement.
The reconstruction is performed by an iterative algorithm by following~\cite{tian2014multiplexed}.

\subsection{The LCNF framework}
\label{NF}
\subsubsection{Data acquisition and preparation}
\label{preprocess}

In our study, we investigated different strategies for training our LCNF network using both experimental and simulated datasets. 

The experimental data was obtained from~\cite{xue2019} and was taken on Hela cells fixed in ethanol or formalin. We collected 22 groups of low-resolution measurements ($2560\times2160 $ pixels) on ethanol-fixed Hela cells and 20 groups of measurements on formalin-fixed Hela cells using multiplexed illumination.  The LED patterns used for illumination are described in Section~\ref{exp setup}, which includes two BF and three DF patterns.

To prepare the input for training the LCNF network, we performed the following steps. Firstly, we extracted the central $250\times250$ pixels from the raw low-resolution intensity images. Then, we applied dynamic range correction by clipping the minimum $0.1\%$ and maximum $0.1\%$ pixel values for each measurement, following the approach described in~\cite{xue2019}. This correction helped suppress shot noise and hot pixels.  Next, we used the DPC reconstruction algorithm, as explained in Section~\ref{dpc}), to generate a linear estimation of the phase based on the two BF intensity measurements. Additionally, we normalized the LED intensities by dividing the intensity images by their mean value. We also applied a morphological open operator to estimate and subtract the slow-changing background, following the method described in~\cite{xue2019}. This process effectively eliminated the unwanted background components and improved the accuracy of the subsequent learning process. Finally, we concatenated the preprocessed low-resolution intensity images with the DPC image. 

To obtain ground-truth high-resolution phase images for the experimental data, we applied the following procedure. Firstly, for each standard FPM measurement, we sequentially illuminated 185 LEDs and captured the corresponding low-resolution intensity images. Then, we employed the model-based FPM reconstruction algorithm, detailed in Section~\ref{fpm}, to reconstruct the phase of the central $250\times250$-pixel region and produce a high-resolution phase image of $1500\times1500$ pixels. Next, we applied a phase unwrapping algorithm~\cite{ghiglia1994robust} to unwrap the reconstructed high-resolution phase image. Furthermore, we addressed the slow-varying background component present in the reconstructed high-resolution phase image by utilizing a morphological open operator with a kernel size of 50. This step removed the slowly changing background, enhancing the clarity and quality of the phase image. To normalize the range of values in the high-resolution phase image, we clipped the phase range within [0, 12] for the Hela cells fixed in both ethanol and formalin. Subsequently, we divided the phase images by this clipping threshold, resulting in a normalized range of values within [0, 1]. Finally, we paired the preprocessed low-resolution input images with the normalized high-resolution reconstructions, which served as the training data for our neural network.

For the simulated dataset, we utilized a portion of the high-resolution DIV2K dataset~\cite{agustsson2017ntire} from the NTRE 2017 challenge~\cite{timofte2017ntire} as our ground-truth phase images. The dataset consisted of 900 cropped natural images, each with a size of $600\times600$ pixels. Since the natural images have different histogram and spectral distributions compared to the biological cell images (see Fig.~\ref{s2}), we performed a preprocessing procedure on these images. The preprocessing involved removing the slowly varying background using an open operator with a kernel size of 20. Then, we applied a maximum value threshold of 0.6 to crop the image values and normalized the cropped images to the range $[0, 1]$ by dividing them by this threshold. To ensure consistency between the simulated dataset and the experimental Hela cells fixed in ethanol (here, we only utilize the data for the Hela cells fixed in ethanol since it contains more frequency content), we took steps to match the power spectrum density (PSD) of the simulated dataset with the experimental data. This involved multiplying the spectrum of each simulated data with a correction map, whose value at a specific frequency is determined by the ratio between the square root of the PSDs of the experimental and the simulated dataset. We then normalized the spectrum-corrected image by dividing it by its maximum value. The resulting normalized and spectrum-corrected high-resolution images were used as ground truth for our network. The effect of the preprocessing procedure for the natural images can be observed in Fig.~\ref{s2}. 

To generate the un-normalized object phase, we multiplied the normalized and spectrum-corrected high-resolution images by a factor of 9 and then subtracted 2.5, resulting in a phase range of $[-2.5, 6.5]$. This range corresponds closely to the predominant distribution observed in the histogram of the experimental ethanol-fixed Hela cell dataset and also balances the effect from the large phase values observed in the experimental dataset (for additional information, refer to Figure~\ref{s2}). To simulate the low-resolution intensity images, we used Eq.~\eqref{equ2} as the forward model and downsampled the simulated intensity images to $100\times100$ pixels.
Throughout the simulation process, we make the assumption that our simulated system does not exhibit aberration. As a result, the pupil function $P(\bu)$ in Eq.~\eqref{equ2} is considered an ideal circular low-pass filter, with a value of 1 within the circular region and 0 outside of it. 
We applied the same preprocessing steps used for the experimental dataset to obtain the preprocessed simulated low-resolution intensity images, which served as the input for the network. Finally, we paired the preprocessed high-resolution images with their corresponding low-resolution intensity images to create the training pairs for the network training.

\subsubsection{LCNF network structure}
\label{net structure}
Our LCNF network consists of a CNN-based encoder and an MLP-based decoder. A detailed visual illustration of the network can be found in Fig.~\ref{s1}. 

\textbf{Encoder:} 
We use three CNN-based encoders, denoted as $\{e_1, e_2, e_3\}$ to independently encode three different types of input: BF, DF, and DPC images. Each encoder follows a deep residual network structure similar to~\cite{lim2017enhanced}. The encoders take specific image types as input and initially extract spatial features using a convolutional layer. The number of input channels for the first convolutional layer varies according to the number of input images: 2 for two BF images, 3 for three DF images, and 1 for the DPC image. The output channels for the first convolutional layer are fixed at 128 for all encoders.

After the initial convolutional layer, we employ 32 residual blocks to further extract spatial feature maps. Each residual block consists of two convolutional layers 128 input and output channels, a ReLU activation layer, and a multiplication layer with a  factor of 1. Skip connections are incorporated in the residual blocks, where feature maps are added together. The spatial features extracted by the residual blocks are then passed through an output convolutional layer with 128 input and output channels. Finally, these features are added with the features maps provided by the initial convolutional layer with a long skip connection. All convolutional layers use $3\times3$ convolutional kernels.

Once the feature maps are extracted from the input, they are concatenated to form the encoded latent space representation $\Phi\in\mathbb{R}^{H\times W\times D}$ of the image, where $H$ and $W$ represent the pixel resolution along the $x$ and $y$ axes, respectively, while D represents the number of concatenated channels. The $H$ and $W$ dimensions remain the same as the input low-resolution measurements, as we do not include pooling or upsampling layers in our encoder networks. The network structure of the encoders is visually depicted in Fig.~\ref{s1}: Encoder.

\textbf{Decoder:} To represent a high-resolution image in a continuous representation, we employ the LIIF approach~\cite{chen2021learning}, which represents an object in the encoded latent space and utilizes an MLP as a decoding function to decode the object from the latent space back to the object domain. In our case, we use a standard 5-layer MLP as the decoder, denoted as $f_{\theta}$. Each layer of the MLP has 256 neurons, and ReLU activation is applied to the first four layers, while the last layer is unactivated. The input dimension of the MLP is 3460, which is obtained by $3$ (number of encoders) $\times$ 128 (feature maps learned by each encoder) $\times$ 9 (feature unfolding) + 2 (dimension of the coordinates) + 2 (cell decoding), where feature unfolding and cell decoding are reconstruction enhancement techniques explained in Section~\ref{enhancement}. The output dimension of the MLP is 1,  representing the predicted phase value at the queried location. The structure of the decoder is illustrated in Fig.~\ref{s1}:~Decoder. 

The decoding function can be expressed as: 
\be
\hat\psi(\bc)=f_{\theta}(\bc, \phi),
\label{eq_s1}
\ee
where $\hat{\psi}(\bc)$ is the decoded physical quantity, such as the phase value in our case, at the queried position $\bc$. The variable $\bc$ represents the 2D coordinates in the continuous image domain, assumed to range in $[-h, h]$ and $[-w, w]$ for the height and the width, respectively. $\phi\in\mathbb{R}^{1\times 1\times D}$ is the selected latent vector from the latent space representation $\Phi$, which is related to the queried position. The decoding function $f_{\theta}(\bc, \phi)$ can be seen as a mapping function $f_{\theta}(\cdot, \phi): \bc \rightarrow \psi(\bc)$ that maps a coordinate to the phase value at the position $\bc$, with the latent vector $\phi$ as conditional parameters. 

It should be noted that the latent space is a low-dimensional space with a dimension $H\times W\times D$, where we assign 2D coordinates to each latent vector with the pre-defined sparse grids, as depicted by the gray circles in Fig.~\ref{s1} Encoder. 
However, for a continuous representation, we may need to query arbitrary coordinates that are not on the predefined grids, as shown by the green circle in Fig.~\ref{s1} Encoder. 
Consequently, we cannot obtain the exact latent vector for the queried position $\bc$ since the density of the grid in the latent space is much lower than that of the high-resolution grid ($H'\times W'$) for the same FOV. 
To bypass this issue, we adopt the LIIF approach~\cite{chen2021learning}, which assumes that the latent space is continuous; in addition, each latent vector can represent a local part of the continuous image and is responsible for predicting the signals at the set of coordinates that are closest to itself. 
Accordingly, we reformulate Eq.~\eqref{eq_s1} as:
\be
\hat\psi(\bc)=f_{\theta}(\Delta\bc, \phi),
\label{eq_s2}
\ee
where $\phi$ is the selected latent vector for coordinate $\bc$, determined by the nearest latent vector based on the Euclidean distance. Here, $\Delta\bc=\bc-v$ and $v$ represent the actual coordinate of the selected latent vector $\phi$, respectively.
Taking Fig.~\ref{s1} Encoder as an example, the bottom-left gray circle represents the selected latent vector, and $v$ denotes the coordinate of this chosen latent vector.

In summary, our network utilizes CNN-based encoders to encode the measurements into a low-dimensional latent space representation, where coordinates are assigned to latent vectors using predefined sparse grids. We can then query the phase value at arbitrary coordinates and use the MLP decoder to decode the physical quantity based on the selected latent vector. The latent space representation, generated by the encoders, adapts to different objects, allowing our decoding function $f_{\theta}(\cdot, \phi)$ to demonstrate robust generalization capabilities compared to traditional NF methods.

\subsubsection{Reconstruction enhancement techniques}
\label{enhancement}
To enhance the information extraction from the latent space and improve the continuity of the reconstruction, we utilize feature unfolding, local ensemble, and cell decoding techniques as described in the LIIF method~\cite{chen2021learning}. 

\textbf{Feature unfolding:}
To capture additional information beyond a single latent vector $\phi$, we employ feature unfolding, which extends $\phi$ to $\hat{\phi}$. Specifically,  $\hat{\phi}$ is obtained by concatenating the $3\times3$ neighboring latent vectors of $\phi$, as illustrated in Fig.~\ref{s3}(a), and is defined as:
\be
\hat{\phi}_{p,q} = \mr{Concat}(\{\phi_{p+l, q+n}\})_{l,n\in\{-1,0,1\}},
\label{eq_s3}
\ee
where Concat represents the concatenation of a set of latent vectors. The indices $p$ and $q$ correspond to the selected latent code $\phi$ that matches the queried coordinate $\bc$ in the latent space. If the queried position is at the image's edge, the latent space $\Phi$ is padded with zero-vectors.

\textbf{Cell decoding:}
We incorporate cell decoding, which takes into account the pixel size information in the decoding function $f_{\theta}$, as illustrated in Fig.~\ref{s3}(b). The updated decoding function is expressed as:
\be
\hat\psi(\bc)=f_{\theta, \mathrm{cell}}([\Delta\bc, c_h, c_w], \hat{\phi}),
\label{eq_s4}
\ee
where the $[c_h, c_w]$ specifies the height and width of the query pixel with the desired pixel size in the reconstruction. The notation $[\Delta\bc, c_h, c_w]$ denotes the concatenation of the coordinate and the pixel size. 
Thus, $f_{\theta, \mathrm{cell}}([\Delta\bc, c_h, c_w], \hat{\phi})$ signifies that the decoding function reconstructs the value with the relative coordinate $\Delta\bc$ and the pixel size $(c_h, c_w)$, conditioned on the ``unfolded'' latent vector~$\hat{\phi}$ at the coordinate $\bc$.

\textbf{Local ensemble}
A concern with Eq.~\eqref{eq_s4} is the discontinuous prediction when the queried coordinate crosses the middle area between two neighboring latent vectors, resulting in a switch between latent codes (i.e. the selection of the nearest latent vector changes). For example, it occurs when the queried coordinate $\bc$ (green dot) crosses the dashed line depicted in Fig.~\ref{s3}(c). Around such coordinates, predictions for two infinitesimally close coordinates are generated based on different latent vectors. Due to imperfections in the learned encoder $E_{\theta_e}$ and decoding function $f_{\theta}$, these borders may exhibit discontinuous patterns.
To address this issue, we employ the local ensemble technique, extending Eq.~\eqref{eq_s4} to:
\be
\hat\psi(\bc)=\sum_{t\in\{00,01,10,11\}} \frac{S_t}{S}\cdot f_{\theta, \mathrm{cell}}([\Delta\bc_t, c_h, c_w], \hat{\phi}_t),
\label{eq_s5}
\ee
where $\hat{\phi}_t~(t\in\{00,01,10,11\})$ represents the four nearest latent vectors (top-left, top-right, bottom-left, bottom-right) based on the queried coordinate, $\Delta\bc_t$ denotes the relative coordinate between the queried coordinate and the selected latent vector, and $S_t$ indicates the area of the rectangle between the queried coordinate and the coordinate of latent vector diagonal to the selected latent vector, as shown in Fig.~\ref{s3}(c). The weights $S_t$ are normalized by $S=\sum_{t} S_t$. Moreover, the latent space representation $\Phi$ is mirror-padded outside the edge, allowing the above formula to work for coordinates near the image borders.

\subsubsection{Network training}
\label{net train}
\textbf{Implementation details:}
To train our network, we follow the procedure outlined in Section~\ref{preprocess} to prepare the training data, which consists of paired input images and the corresponding ground-truth phase images.  During each training step, we randomly crop a smaller patch of size $48\times48$ pixels from the input images. Recall that the size of the raw input measurements differs for the experimental and simulated datasets, with dimensions of $250\times250$ pixels and $100\times100$ pixels respectively.

We encode the input using three encoders, as described in detail in Section~\ref{net structure}, resulting in a latent space representation $\Phi$ with dimensions of $H=48, W=48, D=384$.

Subsequently, we assign 2D coordinates to each latent vector $\phi\in\mathbb{R}^{1\times 1\times 384}$, which is defined on a sparse grid with the same grid density of $48\times48$ as the input. The height and width range of the latent space is set as $[-H, H]$ and $[-W, W]$ respectively, resulting in a distance of 2 between neighboring latent vectors. 

The high-resolution ground-truth phase images, with an original pixel resolution of $1500\times1500$ pixels for the experimental dataset and $600\times600$ pixels for the simulated dataset, are correspondingly cropped into $288\times288$-pixel patches to match the same FOV as the input images.  This scaling indicates that our ground-truth high-resolution phase image has a pixel resolution $6\times$ higher than that of the input in both the $x$ and $y$ directions.

Similar to the assignment of 2D coordinates in the latent space, we assign 2D coordinates to the cropped high-resolution image within the height and width range of $[-H, H]$ and $[-W, W]$ respectively. The grid density is increased to $288\times288$, and the distance between adjacent pixels is reduced to 1/3. This coordinate assignment ensures positional consistency across the measurement domain, latent space, and reconstruction domain, assuming that the information within a 2D image is inherently positionally dependent and the information at a given position is preserved across different domains.

Next, we randomly select 2304 pixels from the high-resolution image patch as the ground-truth phase values by randomly picking the coordinates defined in the high-resolution grid. These coordinates are also used to select the corresponding latent vectors $\phi$ from $\Phi$, as described in Section.~\ref{net structure}. The selected latent vectors and relative coordinates are concatenated and input into the MLP. We further employ the reconstruction enhancement techniques detailed in Section~\ref{enhancement}. 

The output of the MLP is the predicted phase value at the queried position $\bc$, and we train our network by comparing this prediction with the ground truth using the $L_1$ norm, as shown in Eq.~\eqref{equation1}.

It is important to note that during the training stage, we define grids for the high-resolution ground-truth image and query the high-resolution image at these predefined coordinates. 
However, after training, we no longer need to query points at predefined grids and can freely query phase values at any coordinates since our MLP can effectively represent the object in a continuous manner.

We utilize the PyTorch framework for training our network. The Adam optimizer is employed, with an initial learning rate of $1\times10^{-4}$. To adaptively adjust the learning rate during training, we use the \texttt{ReduceLROnPlateau} method in PyTorch. This method reduces the learning rate by a factor of 0.2 when the loss function fails to improve. During training, a batch size of 5 is used. 

\hfill \break
\noindent\textbf{Training with different datasets:}
To comprehensively evaluate the generalization capability of our LCNF framework, in total, we explored three different training strategies and trained five different networks using different datasets, as detailed below.

\begin{description}
\item[Training with the full experimental dataset.] In this case, we trained two networks using two different experimentally captured Hela cell datasets, denoted as \textbf{Network$_{\text{Hela-E(18)}}$} and \textbf{Network$_{\text{Hela-F(16)}}$}. 

For the first Hela(E) dataset, we gathered 22 groups of images for Hela cells fixed in ethanol. \textbf{Network$_{\text{Hela-E(18)}}$} was trained using 18 paired datasets, validated using 2 paired datasets, and tested using 2 paired datasets.

For the second Hela(F) dataset, we captured 20 groups of Hela cells fixed in formalin. \textbf{Network$_{\text{Hela-F(16)}}$} was trained using 16 paired datasets, validated using 2 paired datasets, and tested using 2 paired datasets.

\item[Training with a single pair of experimental dataset.] To further evaluate the generalization capability of our network, we conducted training of two networks using only a single training image pair from two different cell types, denoted as \textbf{Network$_{\text{Hela-E(1)}}$} and \textbf{Network$_{\text{Hela-F(1)}}$}. The remaining images were designated as the test set. This approach allows us to assess the network's performance when trained on extremely limited data, providing insights into its generalization ability and the capability to reduce the complexity of acquiring experimental training data.

\item[Training with the simulated natural images dataset.] In addition to the experimental dataset, we also trained another network using only simulated datasets on natural images, denoted as \textbf{Network$_{\text{Simulate}}$}. The data preparation is described in Section~\ref{preprocess}. For this purpose, we utilized a total of 800 paired images for training, with 50 paired images for validation and another 50 paired images for testing. This simulated dataset allows us to assess the performance of our network in the absence of an experimental training dataset, providing insights into its ability to generalize from simulation to experiment.
\end{description}

For all three training scenarios, the network typically converged at around 500 epochs. 
The training duration varied depending on the dataset. Training the network with a single pair of the experimental dataset took approximately 3 hours while training with the full experimental dataset and the simulated dataset took approximately 24 hours to converge using an NVIDIA Tesla P100 GPU on the Boston University Shared Computing Cluster.

\subsubsection{Network inference}
Upon completion of network training, we can reconstruct high-resolution phase images using a continuous local conditional neural field representation. To perform network inference, we provide the preprocessed measurements of the desired FOV as input and configure the pixel resolution for the resulting reconstructed image.
During the inference process, the network assigns coordinates to each pixel, as described in Section~\ref{net train}, and predicts the corresponding phase value for each queried position.

In contrast to previous NF frameworks~\cite{liu2022recovery} that require a consistent number of input coordinates, resulting in a smaller FOV when aiming for a higher pixel resolution image, our approach maintains the same FOV while increasing the number of queried coordinates. We achieve this by employing varying grid densities for reconstruction.  Notably, the prediction process for a $1500\times1500$-pixel high-resolution image takes approximately 25 seconds, resulting in an average rate of approximately $1\times10^{-5}$ seconds per pixel on a computer with an NVIDIA Quadro RTX8000 GPU.

For the reconstruction of the wide-FOV phase image, we performed inference with a $6\times$ denser grid sampling compared to the raw measurement. We first divided our measurement into a series of small patches with $250\times250$ pixels each. Next, we reconstructed each patch individually, resulting in high-resolution phase images with dimensions of $1500\times1500$ pixels. To create the final wide-FOV reconstruction image, we employed the alpha blending algorithm~\cite{xue2019} to stitch together the individual reconstructions, forming a high-resolution phase image with a diameter of 12960 pixels. It is worth noting that our LCNF network is inherently capable of directly inferring the entire FOV image without requiring any stitching process. However, due to the limitation of GPU memory (48 GB) on our computer, we utilized this patch-wise inference method since direct inference of the entire FOV would exceed the available memory.

\section*{Acknowledgements} 
The authors acknowledge Boston University Shared Computing Cluster for proving the computational resources.
The work is funded by National Science Foundation (1846784).

\section*{Data availability}
The neural network and the data set used in this work are available at \url{https://github.com/bu-cisl/LCNF}.

\section*{Conflict of interest}
The authors declare no competing interests.

\bibliography{cNF}
\bibliographystyle{ieeetr}

\clearpage
\newpage

\renewcommand{\thefigure}{S\arabic{figure}}
\setcounter{figure}{0}

\pagenumbering{arabic} 

\centering
\section*{{\normalfont Supplementary information for:}\\Local Conditional Neural Fields for Versatile and Generalizable Large-Scale Reconstructions in Computational Imaging}

{Hao Wang}$^{1}$, {Jiabei Zhu}$^{1}$, {Yunzhe Li}$^{1,\dagger}$, {Qianwan Yang}$^{1}$, Lei Tian$^{1,2,*}$
\\

[1] Department of Electrical and Computer Engineering, Boston University, Boston, MA 02215, USA.
\\

[2] Department of Biomedical Engineering, Boston University, Boston, MA 02215, USA.
\\

[$\dagger$]{Current address: Department of Electrical Engineering \& Computer Sciences, University of California, Berkeley, CA 94720, USA.}
\\
* Correspondence: leitian@bu.edu

\newpage
\label{SI}

\begin{figure}[h]
\centering
\includegraphics[width=0.95\linewidth]{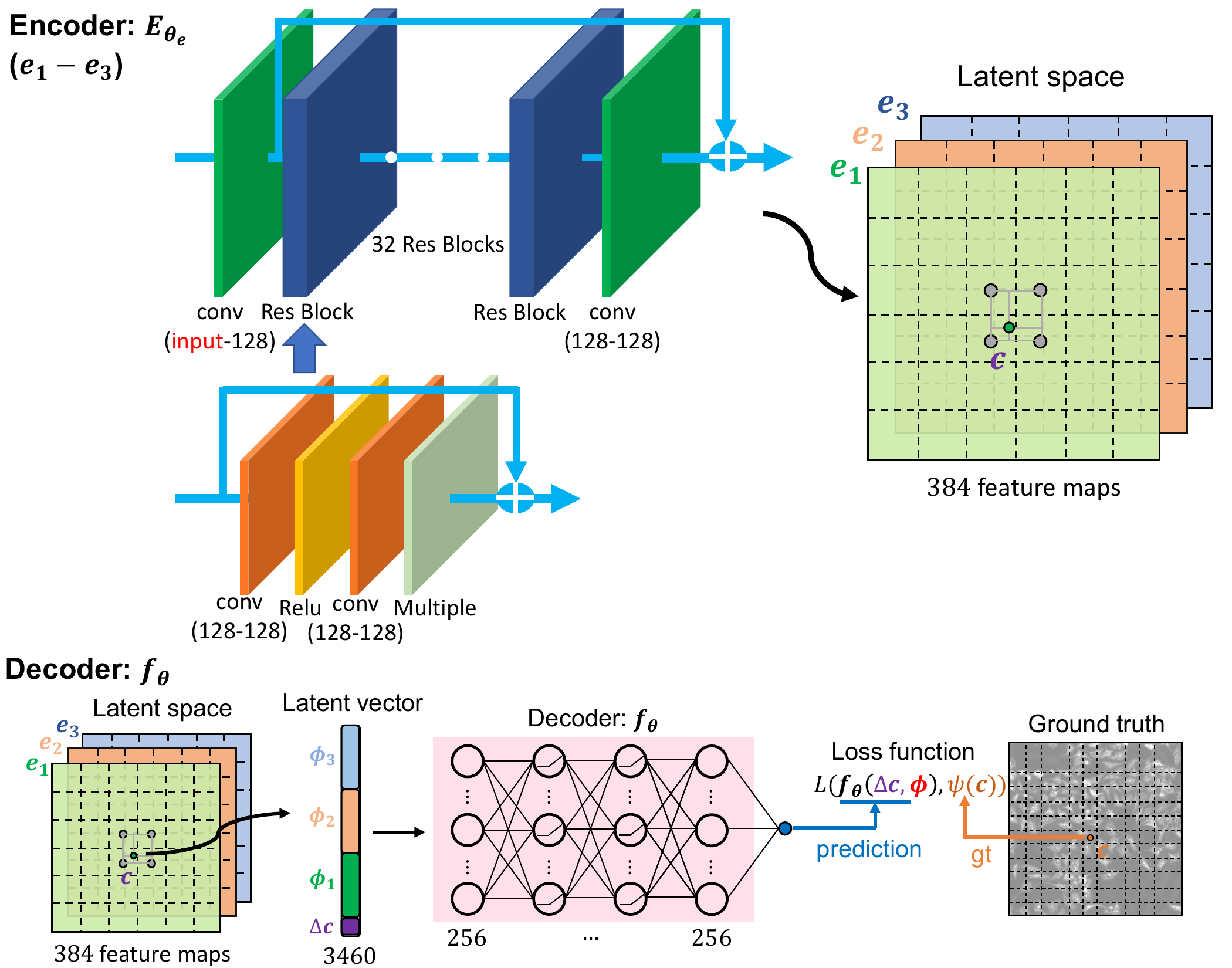}
\caption{Network structure. The LCNF network structure in our framework consists of a CNN-based encoder and an MLP-based decoder.
\textbf{Encoder:} The Encoder module includes three CNNs ($e_1$,$e_2$,$e_3$). Each encoder consists of an input convolutional layer, 32 residual blocks, and an output convolutional layer. The input channel of the input convolutional layer depends on the number of input images, denoted as ``input'' in the figure. For two brightfield intensity images, it has 2 channels. For three darkfield intensity images, it has 3 channels. And for the differential phase contrast image, it has 1 channel. The output channel for all three encoders is set to 128. Each residual block involves two convolutional layers (input channel: 128, output channel: 128), followed by a ReLU activation layer and a multiplication layer with a multiplication factor of 1. Each residual block utilizes a skip connection that adds the feature maps. The spatial features extracted by the residual blocks are then passed through the output convolutional layer, which has 128 input channels and 128 output channels. Finally, features extracted by the output convolutional layer are added with the feature maps provided by the input convolutional layer by a skip connection. All convolutional layers use $3\times3$ convolutional kernels. The feature maps generated by the three encoders are concatenated together to form the encoded latent space representation.
\textbf{Decoder:} To generate a high-resolution image at the queried position based on the encoded latent vectors, we employ a 5-layer MLP with ReLU activation and hidden dimensions of 256 as the decoding function $f_{\theta}$. The latent vector $\phi$ associated with the relative coordinates $\Delta \bc$ of the queried position $\bc$ (green dot) is constructed by concatenating the latent vectors generated by the three encoders: $\phi_1$, $\phi_2$, and $\phi_3$. The input dimension of the MLP is 3460, calculated by $3$ (number of encoders) $\times$ 128 (feature maps learned by each encoder) $\times$ 9 (feature unfolding) + 2 (dimension of the coordinates) + 2 (cell decoding). The output dimension of the MLP is 1, representing the predicted phase value at the queried position. During training, we compared the MLP prediction $f_{\theta}(\Delta \bc,\phi)$ with the ground truth $\psi(\bc)$ to train our network.
}
\label{s1}
\end{figure}

\begin{figure}[h]
\centering
\includegraphics[width=1\linewidth]{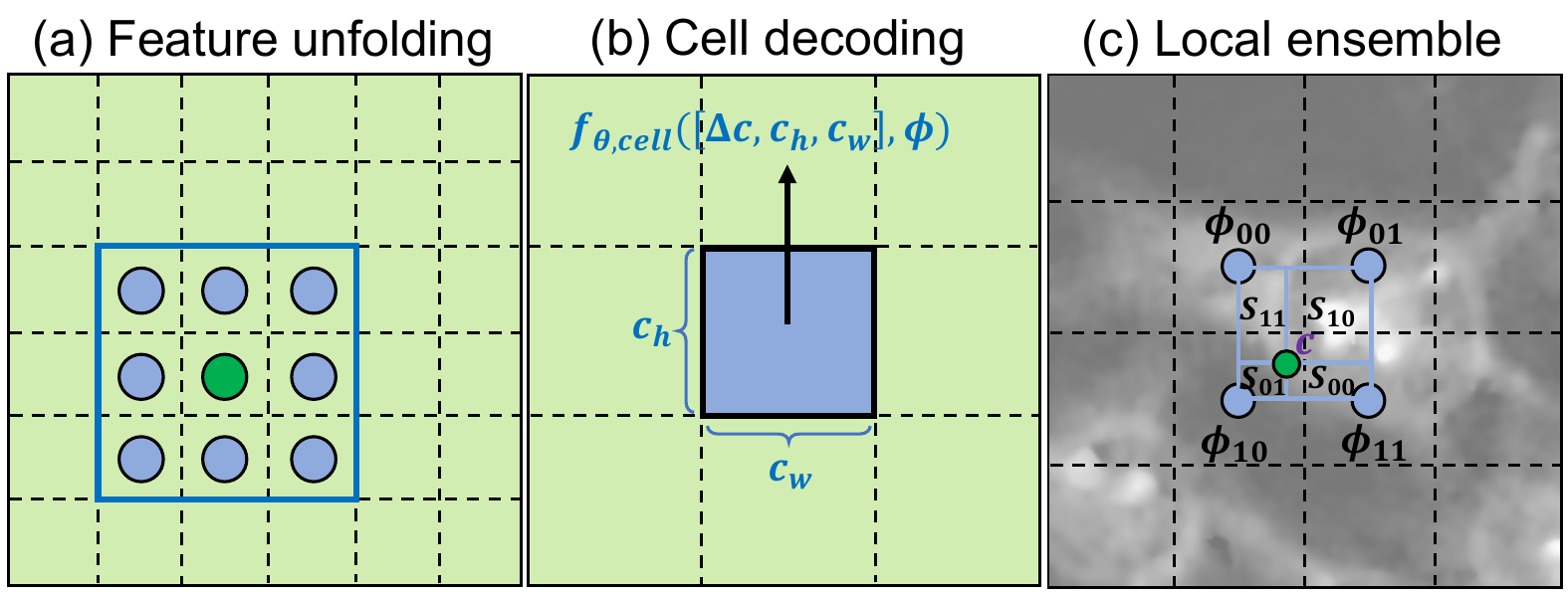}
\caption{Reconstruction enhancement techniques. (a) Feature unfolding. The $3\times3$ neighborhood latent vectors (blue dots) surrounding the selected latent vector (green dot) are concatenated together to provide enriched information. (b) Cell decoding. The pixel size $(c_h, c_w)$ are concatenated with the coordinates to improve the reconstruction. (c) Local ensemble. The local ensemble is used to enhance the continuity of the reconstruction. We utilize the nearest four latent vectors (blue dots, $\{\phi_{00},\phi_{01},\phi_{10},\phi_{11}\}$) to make predictions. These four predictions are then merged by voting with normalized confidences, which are proportional to the area of the rectangle formed between the query point and its selected latent vector’s diagonal counterpart. 
}
\label{s3}
\end{figure}

\begin{figure}[h]
\centering
\includegraphics[width=0.95\linewidth]{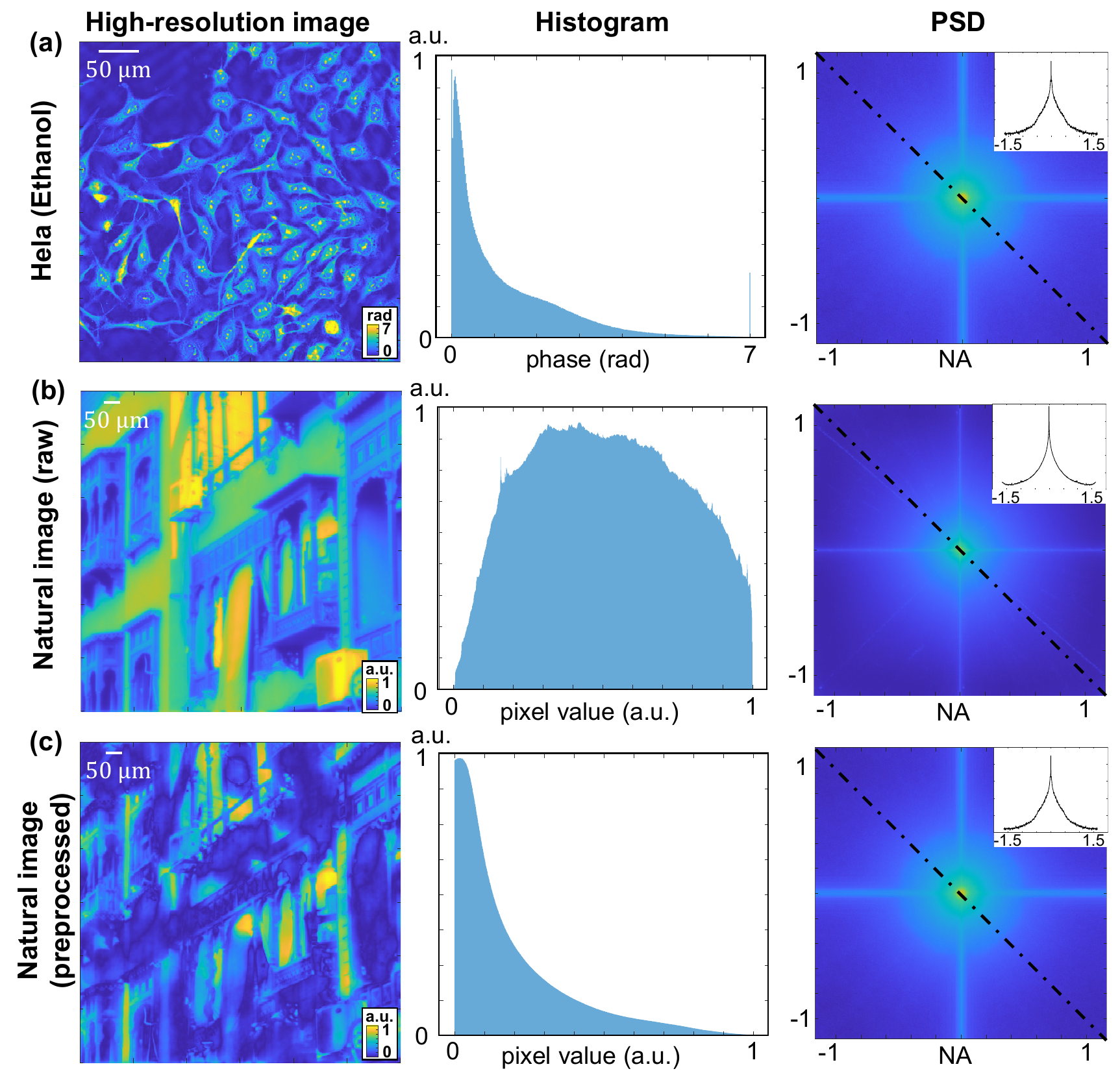}
\caption{Preprocessing of natural images. To align the statistical characteristics of natural images with the experimental cell images, a preprocessing step is performed as described in Section~\ref{preprocess}. The figure presents the effects of this preprocessing step. (a) Model-based reconstruction of ethanol-fixed Hela cells. First column: an example of preprocessed high-resolution model-based FPM reconstruction. Second column: the histogram of the phase values derived from 22 preprocessed high-resolution Hela cell images. Third column: the PSD of the 22 preprocessed high-resolution reconstructions. The figure demonstrates that the phase values are predominantly distributed within the range of [0, 7]. However, a small fraction, approximately $0.5\%$ of pixels lie outside this range. To capture valuable information and eliminate noisy pixels, the phase of the preprocessed experimental dataset is set within the range of [0, 12].
(b) Raw natural images in the DIV2K dataset. First column: an example high-resolution natural image in DIV2K dataset. Second column: the histogram of the pixel values obtained from 900 high-resolution natural images. Third column: the PSD of the 900 high-resolution natural images.
(c) Preprocessed natural images. First column: an example background-removed and spectrum-corrected natural image.
Second column: the histogram of the pixel values derived from 900 preprocessed natural images. We experimentally linearly match the pixel value to the phase range as [0, 9] since it aligns with the predominant distribution of the experimental dataset and also covers possible phase values larger than 7. Third column: the PSD of the 900 preprocessed images.
The profiles of the PSD images along the diagonal (indicated by the dashed black lines) are shown in the top right corner of each image.}
\label{s2}
\end{figure}

\begin{figure}[h]
\centering
\includegraphics[width=1\linewidth]{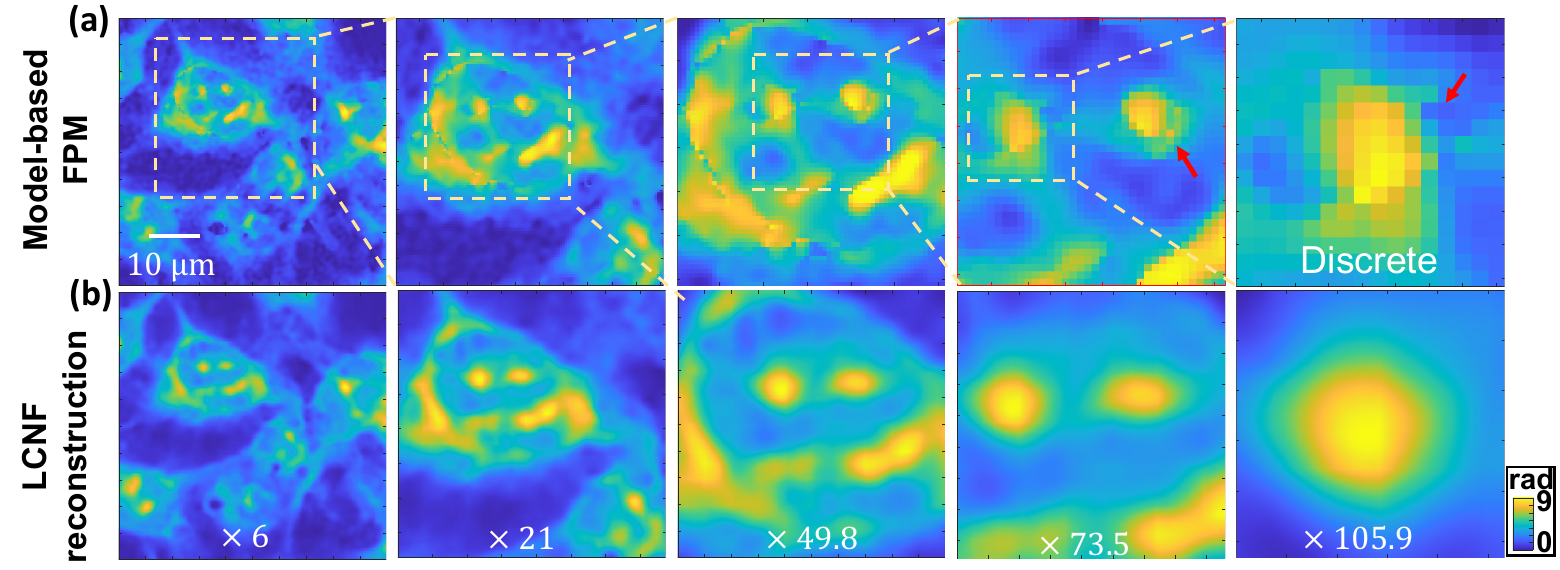}
\caption{The LCNF learns continuous-domain representation and can reconstruct phase maps on an arbitrary pixel grid (illustration with $6×\times, 21\times, 49.8\times, 73.5\times$, and $105.9\times$ upsampling compared with the raw measurement image). (a) Model-based FPM reconstruction with zoom-in areas. Reconstruction artifacts are noted by red arrows. (b) LCNF continuous reconstruction of high-resolution phase image. } 
\label{s8}
\end{figure}

\begin{figure}[h]
\centering
\includegraphics[width=1\linewidth]{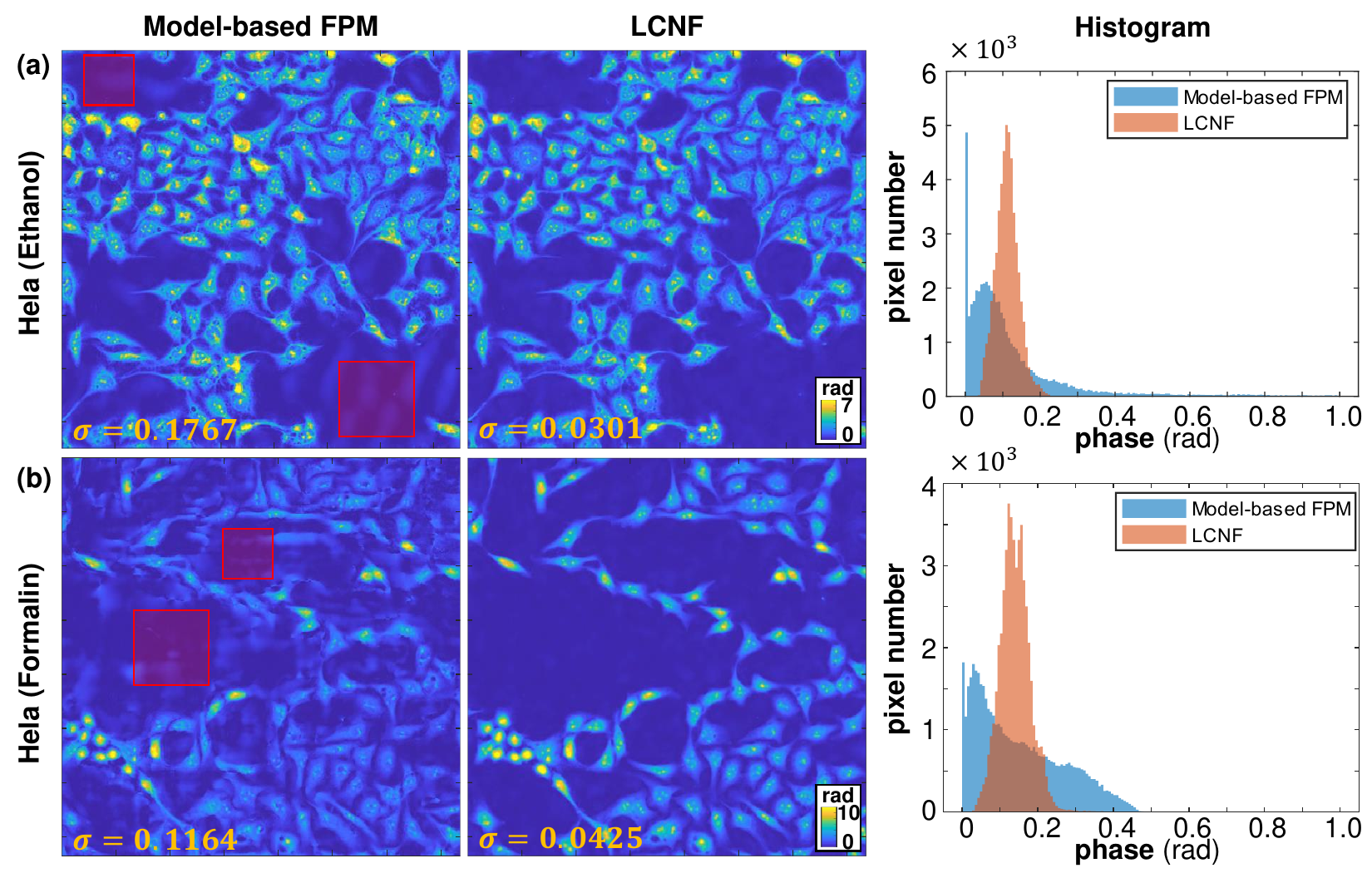}
\caption{Quantitative evaluation of LCNF artifact removal. The red boxes indicate the selected blank areas in the model-based FPM reconstruction and LCNF reconstruction of cells fixed in (a) ethanol and (b) formalin. The histograms of phase values in these blank areas are displayed in the rightmost column. The standard deviation ($\sigma$) of phase values in the selected subareas, indicated at the bottom of the reconstructions, is also calculated. The figure demonstrates that our LCNF effectively eliminates fluctuating artifacts in the background.}
\label{s4}
\end{figure}

\begin{figure}[h]
\centering
\includegraphics[width=1\linewidth]{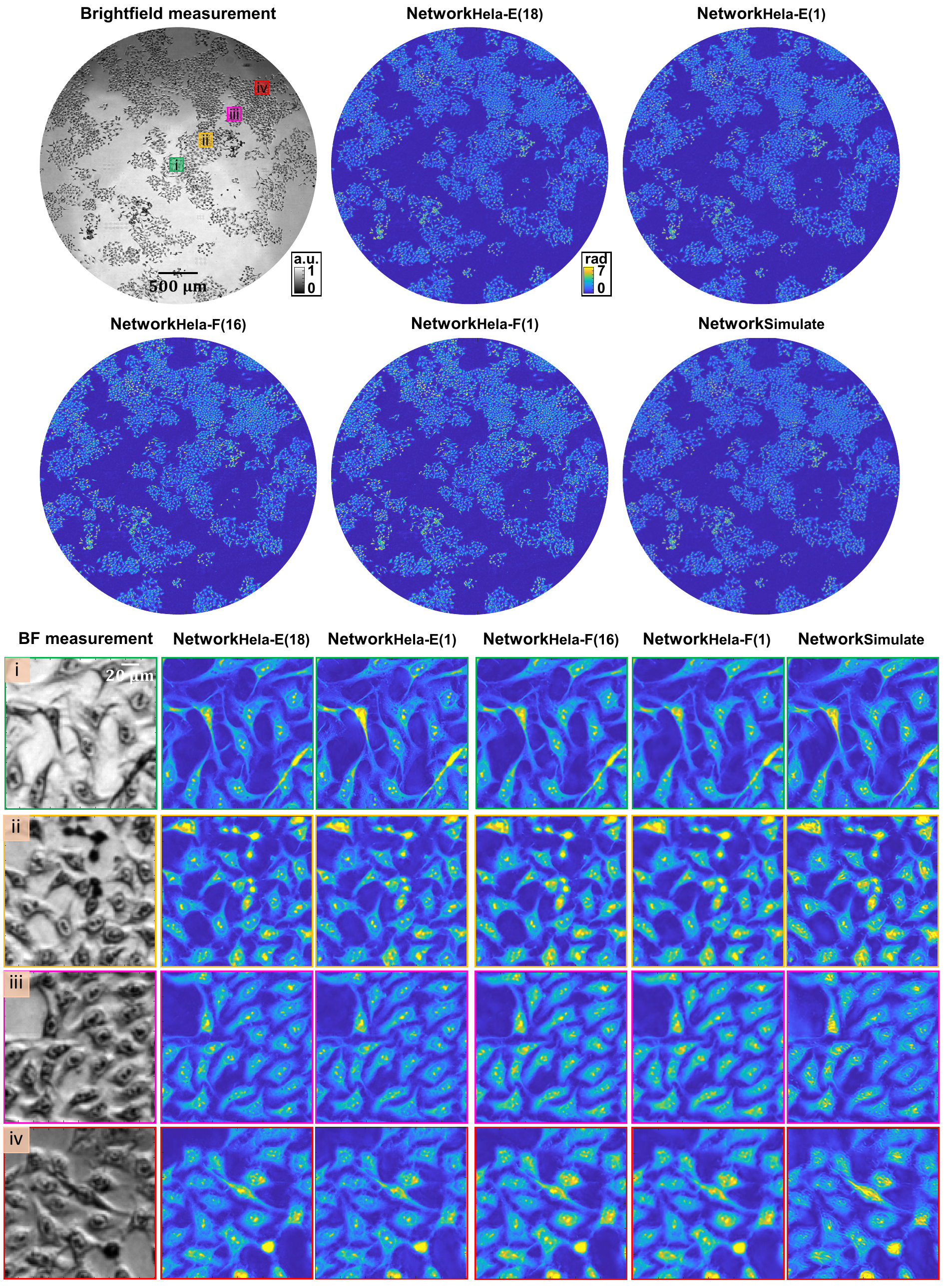}
\caption{Wide-FOV high-resolution reconstructions of Hela cells fixed in ethanol using LCNF networks trained with different datasets.}
\label{s5}
\end{figure}

\begin{figure}[!t]
\centering
\includegraphics[width=1\linewidth]{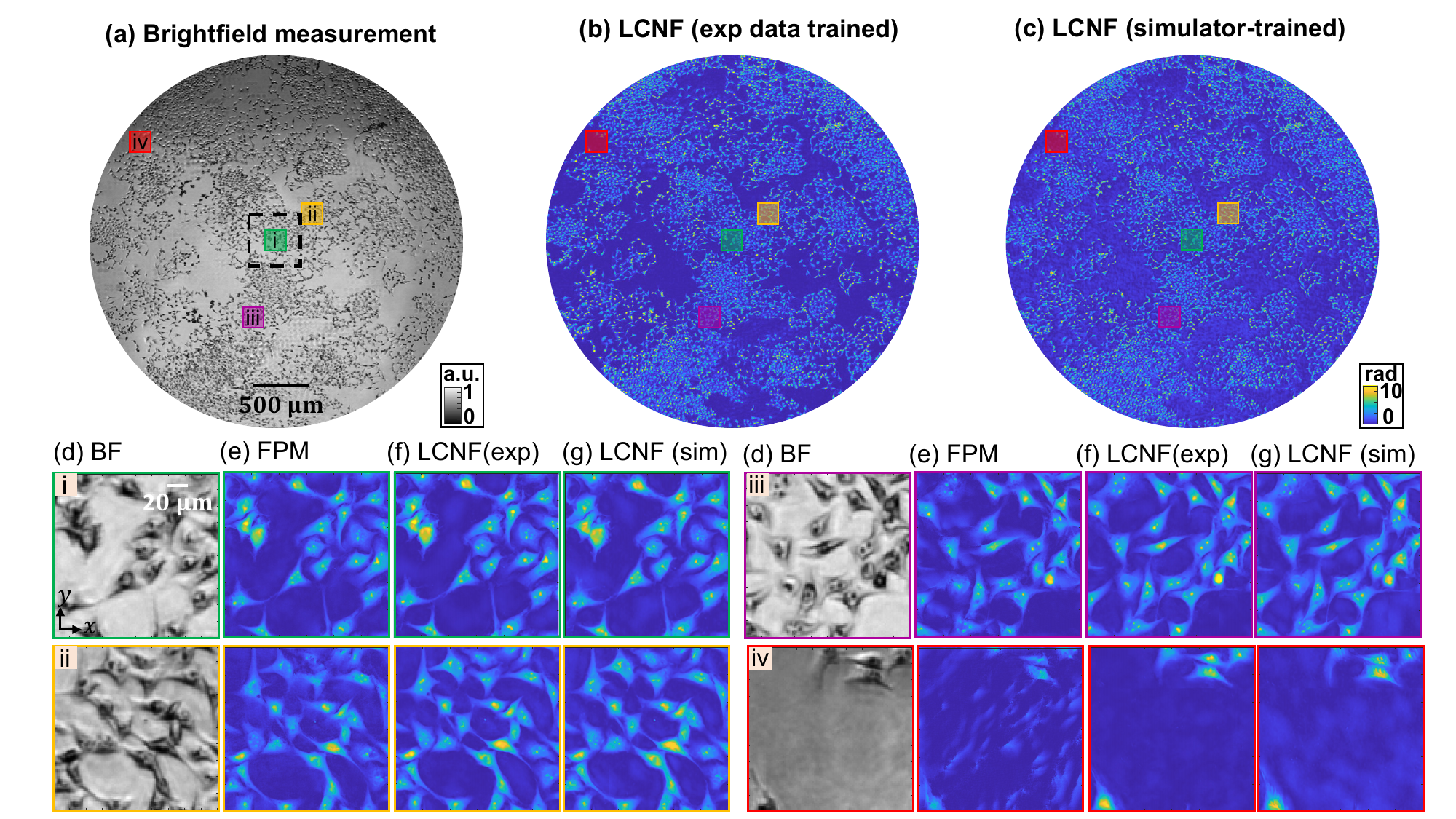}
\caption{Wide-FOV high-resolution LCNF reconstruction of Hela cells fixed in formalin. (a) BF intensity image captured over a 2160-pixel diameter (3.51 mm) FOV. Wide-FOV LCNF reconstruction by (b) \textbf{Network$_{\text{Hela-F(16)}}$} and (c) \textbf{Network$_{\text{Simulate}}$}. (d-g) Selected subareas extracted from the central to the edge of the FOV, identified as i-iv and enclosed within different colored boxes.  (d) BF intensity image. (e) model-based FPM reconstruction. (f) \textbf{Network$_{\text{Hela-F(16)}}$} reconstruction trained with the experimental dataset. The experimental dataset used for training the LCNF network is obtained from the central region, indicated by the dashed black square. (g) \textbf{Network$_{\text{Simulate}}$} reconstruction trained with the simulated dataset. LCNF can successfully reconstruct full-FOV high-resolution images and eliminate the artifacts introduced by the model-based reconstruction, as shown in iv.}
\label{s7}
\end{figure}

\begin{figure}[h]
\centering
\includegraphics[width=1\linewidth]{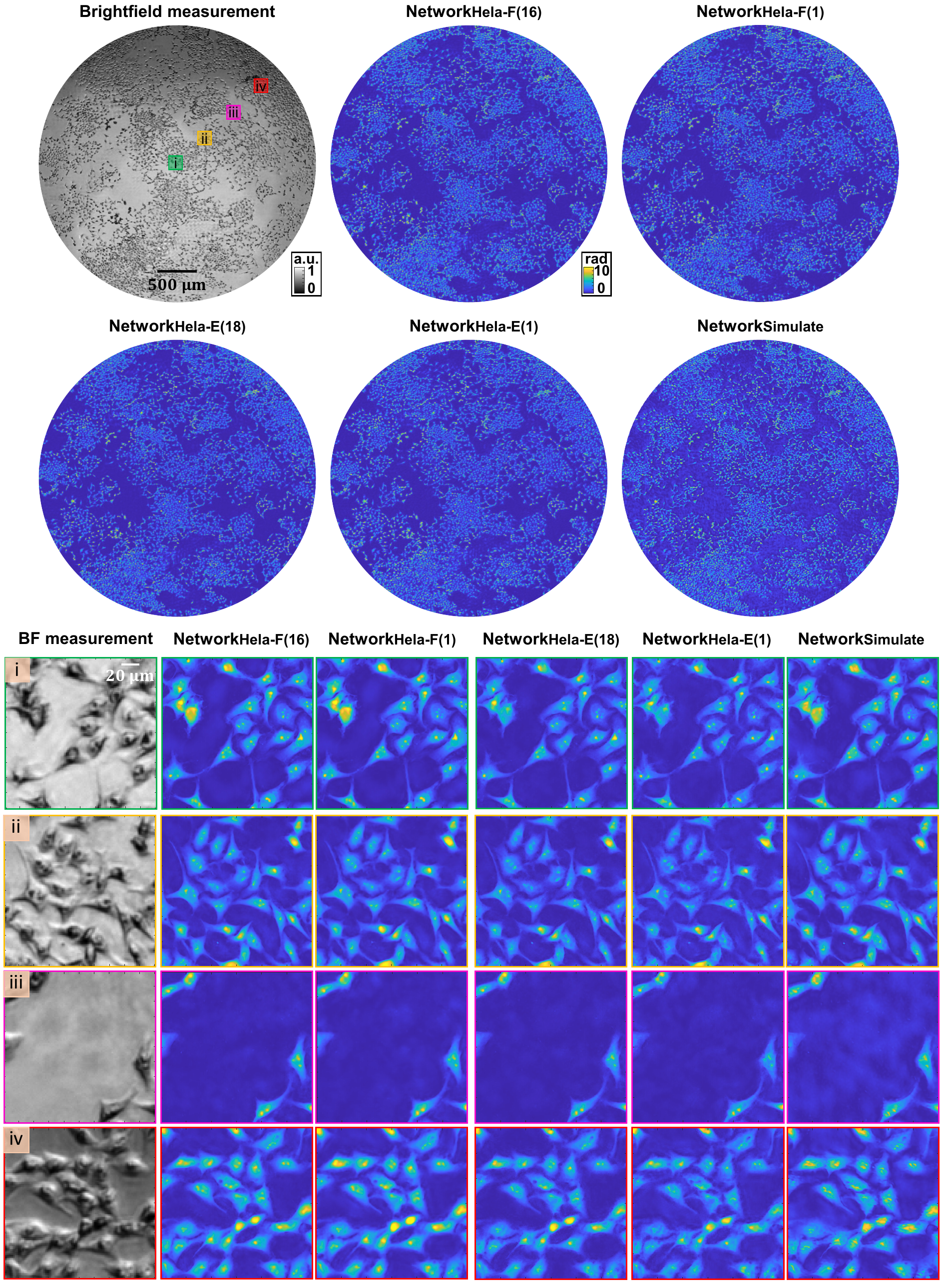}
\caption{Wide-FOV high-resolution reconstructions of Hela cells fixed in formalin using LCNF networks trained with different datasets.}
\label{s6}
\end{figure}

\clearpage

\renewcommand{\thetable}{S\arabic{table}}
\begin{table*}[p]
\centering
\caption{Quantitative metrics for experimental dataset trained LCNF networks. 
In this context, ``Hela (E)'' and ``Hela (F)'' refers to fixed Hela cells preserved in ethanol (E) and formalin (F), respectively.
Here, we consider the model-based FPM reconstruction as the ground truth for assessing the performance of the networks.
	}
\begin{tabular}{ p{1.5cm} p{3cm} p{1.6cm} p{2cm} p{1.6cm} p{1.6cm}}
 \hline
 \textbf{Dataset}   & \textbf{Method}    &\textbf{MSE} &   \textbf{PSNR(dB)}   &\textbf{SSIM}   &\textbf{FM}\\
 \hline
 Hela (E)   & {Network$_{\text{Hela-E(18)}}$}   & $0.0935$   & $28.3800$   & $0.7560$   & $0.0592$\\
  & {Network$_{\text{Hela-E(1)}}$}   & $0.1023$   & $27.9902$   & $0.7611$   & $0.0584$\\
  & {Network$_{\text{Hela-F(16)}}$}   & $0.1139$   & $27.5250$   & $0.7586$   & $0.0552$\\
  & {Network$_{\text{Hela-F(1)}}$}   & $0.1250$   & $27.1220$   & $0.7517$   & $0.0486$\\  
  & {Model-based FPM}   & -   & -   & -   & $0.0788$\\  
 \hline
 Hela (F)   & {Network$_{\text{Hela-F(16)}}$}   & $0.1136$   & $30.6720$   & $0.8093$   & $0.0430$\\
  & {Network$_{\text{Hela-F(1)}}$}   & $0.1527$   & $29.7469$   & $0.7970$   & $0.0396$\\
  & {Network$_{\text{Hela-E(18)}}$}   & $0.1544$   & $29.2155$   & $0.8007$   & $0.0434$\\
  & {Network$_{\text{Hela-E(1)}}$}   & $0.2650$   & $26.8688$   & $0.7528$   & $0.0467$\\
  & {Model-based FPM}   & -   & -   & -   & $0.0510$\\  
 \hline
\end{tabular}
\label{table1:gen}
\end{table*}

\renewcommand{\thetable}{S\arabic{table}}
\begin{table*}[p]
\centering
\caption{Quantitative metrics for simulated dataset trained LCNF. 
In this context, ``Natural image'', ``Hela (E)'', and ``Hela (F)'' refers to the DIV2K natural image dataset, fixed Hela cells preserved in ethanol (E) and formalin (F), respectively.
For the simulated natural image dataset, we have the ground truth high-resolution images from the DIV2K dataset, and for the experiment dataset, we use model-based FPM reconstruction as the ground truth.}
\begin{tabular}{ p{1.5cm} p{3cm} p{1.6cm} p{2cm} p{1.6cm} p{1.6cm}}
 \hline
 \textbf{Dataset}   & \textbf{Method}    &\textbf{MSE} &   \textbf{PSNR(dB)}   &\textbf{SSIM}   &\textbf{FM}\\
 \hline
  Natural image   & {Network$_{\text{Simulate}}$}   & $0.5821$   & $21.4350$   & $0.8533$   & $0.0058$\\
  & Ground truth   & -   & -   & -   & $0.0059$\\
 \hline
 Hela (E)   & {Network$_{\text{Hela-E(18)}}$}   & $0.1317$   & $28.1207$   & $0.7798$   & $0.0534$\\
  & {Network$_{\text{Simulate}}$}   & $0.3261$   & $24.1839$   & $0.6459$   & $0.0680$\\
  & {Model-based FPM}   & -   & -   & -   & $0.0673$\\  
 \hline
 Hela (F)   & {Network$_{\text{Hela-F(16)}}$}   & $0.0766$   & $30.0294$   & $0.8228$   & $0.0383$\\
  & {Network$_{\text{Simulate}}$}   & $0.2503$   & $24.8894$   & $0.7399$   & $0.0418$\\
  & {Model-based FPM}   & -   & -   & -   & $0.0391$\\  
 \hline
\end{tabular}
\label{table2:sim}
\end{table*}
\clearpage

\clearpage
\renewcommand{\thetable}{S\arabic{table}}
\begin{table*}[p]
\centering
\caption{Quantitative metrics for networks trained with varied datasets (mean $\pm$ standard deviation). 
In this context, ``Hela (E)'' and ``Hela (F)'' refers to fixed Hela cells preserved in ethanol (E) and formalin (F), respectively.
The following metrics were computed by comparing the 100 high-resolution phase images ($1500\times1500$ pixels) predicted by the networks trained with different datasets and the model-based FPM reconstructions. The reconstruction patches were extracted from FOVs beyond those in the training dataset.}
\begin{tabular}{ p{1.4cm} p{2.8cm} p{2.1cm} p{2.3cm} p{2.1cm} p{2.1cm}}
 \hline
 \textbf{Dataset}   & \textbf{Method}    &\textbf{MSE} &   \textbf{PSNR(dB)}   &\textbf{SSIM}   &\textbf{FM}\\
 \hline
 Hela (E)   & {Network$_{\text{Hela-E(18)}}$}   & \small$0.1753\pm0.0606$   & \small$30.0157\pm1.7363$   & \small$0.7999\pm0.0243$   & \small$0.0609\pm0.0107$\\
 & {Network$_{\text{Hela-E(1)}}$}   & \small$0.2233\pm0.0818$   & \small$29.0015\pm1.3947$   & \small$0.7275\pm0.0344$   & \small$0.0573\pm0.0099$\\
 & {Network$_{\text{Hela-F(16)}}$}   & \small$0.2356\pm0.0848$   & \small$28.7574\pm1.8154$   & \small$0.7715\pm0.0231$   & \small$0.0531\pm0.0095$\\
 & {Network$_{\text{Hela-F(1)}}$}   & \small$0.2591\pm0.0872$   & \small$28.3186\pm1.6702$   & \small$0.7516\pm0.0290$   & \small$0.0463\pm0.0079$\\
& {Network$_{\text{Simulate}}$}   & \small$0.3704\pm0.1290$   & \small$26.7218\pm1.6527$   & \small$0.6435\pm0.0310$   & \small$0.0772\pm0.0157$\\
& {Model-based}   & -  & -   & -   & \small$0.0689\pm0.0141$\\
 \hline
 Hela (F)   & {Network$_{\text{Hela-F(16)}}$}   & \small$0.1978\pm0.0599$   & \small$30.9115\pm2.0947$   & \small$0.7817\pm0.0484$   & \small$0.0445\pm0.0061$\\
 & {Network$_{\text{Hela-F(1)}}$}   & \small$0.2322\pm0.0685$   & \small$30.1063\pm1.6490$   & \small$0.7738\pm0.0499$   & \small$0.0419\pm0.0059$\\
 & {Network$_{\text{Hela-E(18)}}$}   & \small$0.2074\pm0.0467$   & \small$30.5551\pm1.2916$   & \small$0.7934\pm0.0469$   & \small$0.0448\pm0.0065$\\
  & {Network$_{\text{Hela-E(1)}}$}   & \small$0.3075\pm0.0917$   & \small$28.9166\pm1.1433$   & \small$0.7252\pm0.0557$   & \small$0.0483\pm0.0066$\\
& {Network$_{\text{Simulate}}$}   & \small$0.3358\pm0.0713$   & \small$28.4408\pm1.4819$   & \small$0.6527\pm0.0598$   & \small$0.0503\pm0.0092$\\
& {Model-based}   & -  & -   & -   & \small$0.0479\pm0.0080$\\
 \hline
\end{tabular}
\label{table3:100sample}
\end{table*}
\clearpage

\end{document}